\documentclass[12pt,notitlepage]{article}
\pdfoutput=1
\usepackage{lineno} %For line numbering
\usepackage{threeparttable} %For tables with captions and notes
\usepackage[pdftex]{graphicx} %for inserting pictures into file, graphicx is loaded by tikz
 
\usepackage{StovStyle}

\begin{document}

%%%%%
%%Frontmatter
%%%%%

\title{Fair Compensation\thanks{I thank Scott Condie, Joey McMurray, Juan Moreno-Ternero,
 Brennan Platt, Karol Szwagrzak, and the junior faculty group at BYU's Economics Department for comments.}}
\author{John E. Stovall\thanks{Department of Economics, Brigham Young University, Provo, UT 84602. \href{mailto:johnestovall@byu.edu}{johnestovall@byu.edu}} \\
 Brigham Young University}
\maketitle

\begin{abstract}
We introduce a novel framework that considers how a firm could fairly compensate its workers. A firm has a group of workers, each of whom has varying productivities over a set of tasks. After assigning workers to tasks, the firm must then decide how to distribute its output to the workers. We first consider three compensation rules and various fairness properties they may satisfy. We show that among efficient and symmetric rules: the Egalitarian rule is the only rule that does not decrease a worker's compensation when every worker becomes weakly more productive (Group Productivity Monotonicity); the Shapley Value rule is the only rule that, for any two workers, equalizes the impact one worker has on the other worker's compensation (Balanced Impact); and the Individual Contribution rule is the only rule that is invariant to the removal of workers and their assigned tasks (Consistency). We introduce other rules and axioms, and relate each rule to each axiom.

\vspace{.2in}
\end{abstract}

%\begin{keyword}
%Fair taxation \sep Equal sacrifice \sep Consistency
%
%\JEL D63 \sep D71 \sep D74
%\end{keyword}

%\linenumbers

\clearpage

%%%%%
%%Main Text
%%%%%

\section{Introduction}

What is a fair way for a firm to compensate its workers? A firm has things to do and people to do them. How productive each worker is at the various tasks is, of course, important information that allows the firm to assign workers to tasks so as to maximize output. However, productivity in the various tasks could also be used to determine compensation for the workers. That is, a worker's compensation could depend on her potential production for all tasks.

For example, consider a car dealership that must assign its salespeople to shifts. Some shifts, such as evenings or weekends, might naturally have more sales than other shifts. In addition, each salesperson might be more or less effective depending on the shift. Assuming the manager knows each worker's productivity for every shift, then she could assign the workers to shifts in a way that maximizes sales. This, on average, would be beneficial for the workers as it would mean there is more money available to distribute as compensation. But how should the manager distribute compensation? After all, the workers assigned to the slower shifts still contributed to the dealership's total sales, and they would have contributed more if they had been assigned to the busier shifts.

We envision a scenario in which a firm has some enumerated set of tasks to perform, with each task leading to an identifiable amount of output for that task from a given worker.\footnote{While we assume that the productivity of each worker for each task is known, we recognize that such knowledge may be difficult to attain. However, firms have an obvious incentive to learn these productivities so that they can maximize output. Alternatively, firms without such knowledge could try to design compensation to meet their goals. See \citet{juarez2020psa} for a model in which an uninformed manager wishes to design a compensation mechanism that induces workers to allocate their time among multiple projects so as to maximize output.} Some other examples might be: a firm assigns salespeople to geographical areas; an investment bank assigns traders to trade in specific assets; a software firm assigns developers to work on various projects; a department chair assigns instructors to teach university courses.

How to compensate its workers in one instance would be useful for the firm, but even more useful would be a general rule that tells the firm how to compensate for any scenario they might face. We study a number of different compensation rules and seek to understand them by exploring what general properties, or axioms, they may satisfy. Our axioms generally capture some notion of fairness, and so this axiomatic approach provides us some justification for using a given compensation rule. A firm may be interested in this because --- even if their main goal in choosing a compensation rule is not fairness per se --- then it could be easier for them to justify a rule to their workers for workplace harmony and retention purposes. 

We consider three main compensation rules. The first is the Egalitarian rule, which equalizes compensation across all workers, no matter what their respective productivities. The second rule is the Shapley Value rule, which assigns compensation to each worker according to their average marginal output across all possible orderings of the workers. The third rule is the Individual Contribution rule, which gives each worker exactly what they produce from their assigned task. 

We show that, among efficient and symmetric rules: the Egalitarian rule is the only rule that does not decrease a worker's compensation when every worker becomes weakly more productive  (\autoref{thm01}); the Shapley Value rule is the only rule that, for any two workers, equalizes the impact one worker has on the other worker's compensation (\autoref{thm03}); and the Individual Contribution rule is the only rule that is invariant to the removal of workers and their assigned tasks (\autoref{thm02}).

We also consider other compensation rules and axioms, and we relate each rule to each axiom. All results are summarized in \autoref{table02}. There are too many to discuss at this time, but most of the surprising negative results are for the Individual Contribution rule. For example, we show that the Individual Contribution rule is not continuous (\autoref{ex02}), may compensate a worker less than another even when that individual has an absolute advantage in all tasks (\autoref{ex04}), may compensate a worker less when their productivity increases (\autoref{ex05}), may simultaneously benefit some workers and harm other workers when new workers are hired (\autoref{ex06}), and may alter compensation if a null task (a task in which every worker has zero productivity) is removed from consideration (\autoref{ex07}). 

In addition, we show that the Shapley Value rule violates a mild additivity property (\autoref{ex09}). This is surprising since the original characterization of the Shapley value in the context of transferable utility coalitional games employed an additivity property \citep{shapley1953avf}.

The framework we develop here is novel, and so there are no directly comparable papers. However, the problem of allocating some finite resource that is generated by a group of agents is pervasive in social choice theory. Most notably, the problem we consider can be mapped into a transferable utility coalitional game. Indeed, our characterization of the Shapley Value rule is borrowed from \citet{myerson1980conference}. However, the additional structure of our model, as well as the lack of coalition formation, means that our model requires a separate analysis from coalitional games. We discuss this point further in \autoref{Sec_Coalitions}.

Other social choice problems concerning fairly allocating a surplus generated by a group of agents include: distributing revenue generated by museum passes \citep{ginsburgh2003tmp}; sharing proceeds from a venture with a hierarchical structure \citep{hougaard2017stp,hougaard2022omo}; and sharing revenue from sports broadcasts \citep{bergantinos2020str}.

On a historical note, this work was inspired by an experiment by \citet{gihleb2020}. The purpose of their paper was to explore the effect of negotiation on compensation. In their experiment, three subjects are assigned to be either a manager or one of two workers. Each worker is then assigned to perform either a high- or low-productivity task. After the tasks are completed, the manager must distribute the group's output to the workers. The non-obviousness of fair payment due to the disparity in production from the two tasks creates an interesting environment in which the two workers can negotiate for compensation. This non-obviousness also led to the present work.

\section{Framework}

We use the following notation. Let $\mathcal{N}$ denote the set of all finite subsets of $\mathbb{N}$. For $N \in \mathcal{N}$, let $|N|$ denote the cardinality of $N$. For $x,y \in \mathbb{R}^N$, we use the vector inequalities $x \geqq y$ if $x_n \geq y_n$ for all $n \in N$, $x \geq y$ if $x \geqq y$ and $x \neq y$, and $x > y$ if $x_n > y_n$ for every $n \in N$. Similar inequalities will apply to matrices. For $x \in \mathbb{R}^N$ and $N' \subset N$, let $x_{N'}$ denote the projection of $x$ onto $\mathbb{R}^{N'}$. For $n \in N$, let $x_{-n}$ denote $x_{N \setminus \{ n \} }$.

\subsection{Definitions}

Let $I \in \mathcal{N}$ be a set of individuals. Let $T \in \mathcal{N}$ be a set of tasks where $|T| \geq |I|$. Each individual will produce some homogeneous and divisible output when performing a task. Let $p_{i}^{t} \in \mathbb{R}_+$ denote the amount of output individual $i$ can produce when assigned task $t$. Let $p = (p_{i}^{t})_{i \in I , t \in T}$ denote the \emph{productivity matrix}. For $i \in I$, let $p_i = (p_{i}^t)_{t \in T}$ be individual $i$'s \emph{productivity profile}, and let $p_{-i}$ denote the productivity matrix that is formed by removing $p_i$ from $p$. For $I' \subset I$, let $p_{I'}$ denote the productivity matrix identical to $p$ but restricted to the individuals in $I'$. In a similar manner, for $t \in T$ and $T' \subset T$ satisfying $|T'| \geq |I|$, we define $p^t$, $p^{-t}$ and $p^{T'}$.

A \emph{compensation problem} is a tuple $(I,T,p)$ where $I,T \in \mathcal{N}$ satisfy $|I| \leq |T|$ and $p \in \mathbb{R}_{+}^{I \times T}$. An \emph{assignment} is a function $a:I \rightarrow T$ that is injective. We interpret $a(i)=t$ to mean that individual $i$ has been assigned task $t$. %\footnote{A natural generalization of an assignment would assign a portion of a task to an individual. I.e. an assignment would be a function $a: I \times T \rightarrow [0,1]$ such that (1) for every $i \in I$, we have $\sum_{t \in T} a(i,t) = 1$, and (2) for every $t \in T$, we have $\sum_{i \in I} a(i,t) \leq 1$. However, a corner solution would always be optimal in this case, implying an assignment as defined above.}
Note that the number of possible assignments for any problem is $\genfrac(){0pt}{1}{|T|}{|I|}$, and thus finite. Let $A(I,T)$ denote the set of all possible assignments for $I$ and $T$, and define
\[
y(I,T,p) \coloneqq \max_{a \in A(I,T)} \sum_{i \in I} p^{a(i)}_i \text{.}
\]
Thus $y(I,T,p)$ is the maximum amount of output possible under the problem $(I,T,p)$. Let $A^*(I,T,p)$ denote the set of optimal assignments for $(I,T,p)$, i.e.
\[
A^*(I,T,p) \coloneqq \arg\max_{a \in A(I,T)} \sum_{i \in I} p^{a(i)}_i \text{.}
\]
Thus $y(I,T,p) = \sum_{i \in I} p^{a(i)}_i$ for every $a \in A^*(I,T,p)$.\footnote{An optimal assignment is not chosen according to comparative advantage. For example, if $p_1 = (2,1)$ and $p_2 = (5,3)$, then individual 1 has a comparative advantage in task 1. Yet the optimal assignment has $a^*(1)=2$.}

A \emph{solution} for the problem $(I,T,p)$ is a vector $x \in \mathbb{R}_+^I$ such that $\sum_{i \in I} x_i \leq y(I,T,p)$. A \emph{compensation rule} is a function that maps compensation problems to solutions. In what follows, we will use $R$ to denote a generic compensation rule.

\subsection{Compensation Problems and Coalitional Games} \label{Sec_Coalitions}

A compensation problem could be viewed as a transferable utility coalitional game. That is, for the compensation problem $(I,T,p)$, define the characteristic function $v(S) = y(S,T,p_S)$ for $S \subset I$. While viewing a compensation problem in this way can be informative, there are two important caveats that make our analysis more than just an application of this literature.

First, a compensation problem contains more information than a characteristic function in a coalitional game because the productivity of each agent in each task is known. Not only does this provide a specific structure as to how the surplus is determined, it also gives information about the contribution of each individual. Since our analysis is centered on fairness of the compensation rule, this is potentially relevant information that can be used to determine such compensation. In addition, because of this added structure, adapting axioms from the coalitional game setting does not always yield similar results in our setting. As noted in the introduction, the Shapley Value rule does not satisfy a mild additivity property (\autoref{ex09}).

Second, strategic considerations are not a factor in our analysis since a compensation problem represents a fixed firm's reality. That is, for the purposes of this normative exercise, individual workers are assumed to be a member of the organization and thus not able to split off to form competing coalitions. In addition, the production technology is assumed to be owned by the firm and thus the value to an individual of leaving a firm is not known from the specification of a compensation problem. For example, for a given problem $(I,T,p)$, individual $i \in I$ has the productivity profile $p_i$, but since the organization is the one that owns the production process, there is no guarantee that $i$ would have the same productivity profile in another firm, or just producing by herself. Indeed, there is no guarantee that $T$ would be the set of tasks available in another firm. Thus $\max_t p_i^t$ should be thought of as individual $i$'s productivity if she were the only worker in the firm, not as being individual $i$'s threat point were she to splinter off and form her own firm. 

Continuing this point, many of the solution concepts in cooperative game theory are not as compelling in this context. For example, the core no longer represents the set of stable divisions since coalitions cannot form and splinter off of the grand coalition.

One other thing to note when viewing a compensation problem as a coalitional game is that in general such games are submodular. See \autoref{Sec_NHfH}.

\section{Main Results}  \label{Sec_Results}

We axiomatically characterize three compensation rules: the Egalitarian, Shapley Value, and Individual Contribution rules.

However, before presenting these results, we introduce two basic axioms which all our rules will satisfy. The first says that total compensation must exhaust all of the output.

\begin{axiom}[Efficiency]
For every $(I,T,p)$, we have $\sum_{i \in I} R_i(I,T, p) = y(I,T,p)$.
\end{axiom}

Efficiency is the basic requirement that there be no waste when compensating the individuals. All of our characterization results will employ this axiom. An example of a rule that would violate Efficiency is a rule that gives everyone zero compensation.

Efficiency is a ubiquitous property. However, in this context, one criticism is that distributing all of the output to the workers leaves nothing for the owner of the firm. This could be addressed in a number of ways. One is to simply include the owner as one of the workers, as well as a task that only the owner can perform (i.e. the other workers have zero productivity in this task). Another way is to modify Efficiency so that the sum of compensation to all workers is equal to some fixed $\alpha \in (0,1]$ share of total output. All of our results would still hold with straightforward modifications of our rules. Finally, if the firm is collectively owned by the workers --- a worker cooperative --- then Efficiency is self-evidently desirable.

The next axiom is a fairness condition that requires a rule to treat identical individuals equally.

\begin{axiom}[Symmetry]
For every $(I,T,p)$ and for every $i,j \in I$, if $p_i = p_j$, then $R_i(I,T,p) = R_j (I,T,p)$.
\end{axiom}

Only the characterization of the Egalitarian rule will employ Symmetry, even though all of our rules will satisfy it. An example of a rule that would violate Symmetry is a dictatorial rule --- i.e. a rule that chooses one individual and always gives them all of the output.

\subsection{The Egalitarian Rule}

The Egalitarian rule divides a firm's output evenly between all of its workers.

\begin{axiom}[Egalitarian Rule]
For any problem $(I,T,p)$ and for any $i \in I$, the Egalitarian rule $E$ assigns to $i$ the compensation
\[
E_i (I,T,p) \coloneqq \frac{y(I,T,p)}{|I|} \text{.}
\]
\end{axiom}

Note that this method of compensation completely ignores the contribution of each individual --- an individual's productivity is not directly used in determining their own compensation. 

The key axiom in our characterization of the Egalitarian rule is a group monotonicity property with respect to the productivity matrix.

\begin{axiom}[Group Productivity Monotonicity]
For every $I,T \in \mathcal{N}$ where $|I| \leq |T|$ and for every $p,\hat{p} \in \mathbb{R}_{+}^{I \times T}$, if $p \geqq \hat{p} $, then $R(I,T,p) \geqq R(I,T,\hat{p})$.
\end{axiom}

Group Productivity Monotonicity can be thought of as a group solidarity property. If each individual in the group becomes weakly more productive, then no individual should be worse off.

\begin{theorem} \label{thm01}
The Egalitarian rule is the only compensation rule satisfying Efficiency, Symmetry, and Group Productivity Monotonicity.
\end{theorem}
\begin{proof}
It is straightforward to show that the Egalitarian rule satisfies the three axioms.

To show the Egalitarian rule is the only such rule, let $R$ be a rule that satisfies Efficiency, Symmetry, and Group Productivity Monotonicity. Fix the problem $(I,T,p)$. If $|I|=1$, then Efficiency implies that $R(I,T,p) = y(I,T,p) = E(I,T,p)$.

So assume $|I| \geq 2$. Fix $a \in A^*(I,T,p)$. Let $\hat{p}$ be the productivity matrix where
\[
\hat{p}_{i}^t = \begin{cases}
p_{i}^t & \text{ if } t = a(i) \\
0 & \text{ otherwise}
\end{cases}
\]
for every $i \in I$. Note that $p \geqq \hat{p}$. Hence, Group Productivity Monotonicity implies $R(I,T,p) \geqq R(I,T,\hat{p})$. Note also that $y(I,T,p) = y(I,T,\hat{p})$ since $a \in A^*(I,T,p)$, $a \in A^*(I,T,\hat{p})$, and $\sum_{i \in I} p_i^{a(i)} = \sum_{i \in I} \hat{p}_i^{a(i)}$. So Efficiency implies $R(I,T,p) = R(I,T,\hat{p})$.

Let $\{i,j\} \subset I$. Let $\tilde{p}$ be the productivity matrix where $\tilde{p}_{i}^{a(j)} = \hat{p}_{j}^{a(j)}$, $\tilde{p}_{j}^{a(i)} = \hat{p}_{i}^{a(i)}$, and $\tilde{p}_{k}^{t} = \hat{p}_{k}^{t}$ otherwise. Similar to the procedure above, we have $\tilde{p} \geqq \hat{p}$ and $y(I,T,\tilde{p}) = y(I,T,\hat{p})$. Hence, Group Productivity Monotonicity and Efficiency imply $R(I,T,\tilde{p}) = R(I,T,\hat{p})$. 

Thus we have $R(I,T,p) = R(I,T,\tilde{p})$. But since $\tilde{p}_i = \tilde{p}_j$, Symmetry implies $R_i(I,T,\tilde{p}) = R_j(I,T,\tilde{p})$, which means we have $R_i(I,T,p) = R_j(I,T,p)$.

Since $i$ and $j$ were chosen arbitrarily in the previous step, this implies $R_i(I,T,p) = R_j(I,T,p)$ for any $i,j \in I$. Efficiency then implies $R_i (I,T,p) = \frac{y(I,T,p)}{|I|}$ for every $i \in I$.
\end{proof}

\autoref*{thm01} demonstrates that Group Productivity Monotonicity is a strong axiom. The strength stems from the following implication: if individual $i$'s productivity stays constant while individual $j$'s productivity increases, then $i$ should not be punished and receive less compensation. This implication does have normative appeal, but there may also be good reasons to give $i$ less compensation in this situation because of how tasks are assigned. Namely, if $j$ becomes more productive, then that may mean that $i$ is assigned a different task and produces less. Thus $j$ becomes more valuable to the group while $i$ becomes less valuable. In \autoref{Sec_Mono}, we introduce a weaker productivity monotonicity property for individuals that does not have the above implication, and is satisfied by many of the other rules we will introduce. Thus, one way to view \autoref*{thm01} is that it demonstrates that if one is to insist on the group solidarity property inherent in Group Productivity Monotonicity as opposed to the weaker individual version, then compensation must be equal.

\subsection{The Shapley Value Rule}

Our next compensation rule uses a well-known method for dividing a surplus.\footnote{To make the definition of the Shapley Value rule well-defined, set $y(\emptyset,T,p) \coloneqq 0$.}

\begin{axiom}[Shapley Value Rule]
For any problem $(I,T,p)$ and for any $i \in I$, the Shapley Value rule $SV$ assigns to $i$ the compensation
\[
SV_i (I,T,p) \coloneqq \sum_{J \subset I \setminus \{i\}} \frac{|J|!(|I| - |J|-1)!}{|I|!} \left[ y(J \cup \{i\},T,p_{J \cup \{i\}}) - y(J,T,p_J) \right] \text{.}
\]
\end{axiom}

Imagine that the individuals in group $I$ are lined up in a queue and added one-by-one to the production of the firm. Each time an individual is added, the individuals are reassigned tasks so as to maximize output. If $J \subset I$ are the individuals in the queue that come before $i$, then $y(J \cup \{i\},T,p_{J \cup \{i\}}) - y(J,T,p_J)$ is $i$'s marginal contribution to the output. The Shapley Value rule compensates an individual by giving them the average of their marginal contribution over all possible orderings of the group $I$.

The key axiom satisfied by the Shapley Value rule is Balanced Impact.

\begin{axiom}[Balanced Impact]
For every $(I,T,p)$ where $|I| \geq 2$, for every $\{ i,j \} \subset I$, we have $R_i(I,T,p) - R_i (I \setminus \{ j \} , T, p_{-j}) = R_j(I,T,p) - R_j (I \setminus \{ i \} , T, p_{-i})$.
\end{axiom}

For the compensation problem $(I,T,p)$ and for $\{ i,j \} \subset I$, the difference $R_i(I,T,p) - R_i (I \setminus \{ j \} , T, p_{-j})$ measures $j$'s impact on $i$'s compensation. Balanced Impact is a fairness condition that insists that the compensation rule equalize the impact two individuals have on each other.

Balanced Impact was first introduced by \citet{myerson1980conference} and used to characterize the Shapley value as a solution for coalitional games. We get a similar result for compensation problems.

\begin{theorem} \label{thm03}
The Shapley Value rule is the only compensation rule satisfying Efficiency and Balanced Impact.
\end{theorem}
\begin{proof}
Showing the Shapley Value rule satisfies Efficiency and Balanced Impact is straightforward, and so is omitted.

To show it is the only such rule, suppose that $R'$ and $R''$ both satisfy Efficiency and Balanced Impact. We show that $R'(I,T,p) = R''(I,T,p)$ for any $(I,T,p)$. We do this by mathematical induction on $|I|$. Note that Efficiency implies that $R'(I,T,p) = R''(I,T,p)$ for every problem $(I,T,p)$ where $|I|=1$. Now suppose that $n \in \mathbb{N}$ is such that for every $(I,T,p)$ where $|I| = n$, we have $R'(I,T,p) = R''(I,T,p)$. Fix problem $(I,T,p)$ such that $|I| = n+1$, and fix $\{ i,j \} \subset I$. By Balanced Impact, we have
\[
R'_i(I,T,p) - R'_j(I,T,p)  = R'_i (I \setminus \{ j \} , T, p_{-j}) - R'_j (I \setminus \{ i \} , T, p_{-i})
\]
and 
\[
R''_i(I,T,p) - R''_j(I,T,p)  = R''_i (I \setminus \{ j \} , T, p_{-j}) - R''_j (I \setminus \{ i \} , T, p_{-i}) \text{.}
\]
Since $| I \setminus \{ j \} | = | I \setminus \{ i \} | = n$, the inductive hypothesis implies 
\[
R'_i (I \setminus \{ j \} , T, p_{-j}) - R'_j (I \setminus \{ i \} , T, p_{-i}) = R''_i (I \setminus \{ j \} , T, p_{-j}) - R''_j (I \setminus \{ i \} , T, p_{-i}) \text{,}
\]
which in turn implies 
\[
R'_i(I,T,p) - R'_j(I,T,p)  = R''_i(I,T,p) - R''_j(I,T,p)  \text{.}
\]
Since $\{ i,j \} \subset I$ was chosen arbitrarily, we thus have $R'(I,T,p) = R''(I,T,p)$. This completes the inductive step.
\end{proof}

The usual critique of Balanced Impact for coalitional games applies here. Namely, it is not clear why one would want the impact two workers impose on each other to be equal. For example, if one worker was highly productive while another was not, then it may seem natural for the highly productive worker to have a larger impact on the other worker.

\subsection{The Individual Contribution Rule}

Our final rule assigns compensation according to how much an individual personally contributes to the overall output.

\begin{axiom}[Individual Contribution Rule] 
For any problem $(I,T,p)$ and for any $i \in I$, the Individual Contribution rule $IC$ assigns to $i$ the compensation
\[
IC_i (I,T,p) \coloneqq \frac{ 1 }{|A^* (I,T,p)|} \sum_{a \in A^* (I,T,p)} p_{i}^{a(i)} \text{.}
\]
\end{axiom}

Note that if $A^*(I,T,p) = \{ a \}$, then $IC_i (I,T,p) = p_{i}^{a(i)}$. That is, each individual is compensated with their individual production. More generally, if $A^*(I,T,p)$ is not a singleton, then individual $i$ gets the average of their productivity under all possible optimal assignments. 

Before introducing the key axiom, we set forth some notation. For problem $(I,T,p)$, assignment $a \in A(I,T)$, and subgroup $I' \subset I$, let $a(I')$ denote the image of $I'$ under $a$. Thus $a(I')$ are all the tasks assigned by $a$ to the individuals in $I'$. Note also that $p_{I'}^{a(I')}$ is the productivity matrix $p$ restricted to the individuals in $I'$ and their assigned tasks under $a$.

\begin{axiom}[Weak Consistency]
For every $(I,T,p)$ where $A^*(I,T,p) = \{a\}$ is a singleton, and for every $I' \subset I$, we have 
\[
R_{I'} (I,T,p) =  R\left(I',a(I'),p_{I'}^{a(I')}\right) \text{.}
\]
\end{axiom}

Weak Consistency is an invariance condition with respect to the removal of individuals and tasks from the compensation problem. Imagine a scenario in which some individuals are removed from the problem $(I,T,p)$, leaving only $I' \subset I$. In addition, some tasks are also removed from the problem. The tasks that remain are exactly the set of tasks that are optimal to assign to $I'$ in the problem $(I,T,p)$, namely $a(I')$. The ``reduced'' compensation problem consisting of just these individuals and just these tasks would be $(I',a(I'),p_{I'}^{a(I')})$. Note that for the individuals in $I'$, there are many things that are constant across the problems $(I,T,p)$ and $(I',a(I'),p_{I'}^{a(I')})$. In particular, relative to the set of tasks that they perform collectively, their productivities are constant. This implies that the optimal assignment of tasks does not change, and thus as a group they produce the same amount of output. Consistency thus imposes the requirement that the compensation each individual in $I'$ receives should be constant across these different problems.

Consistency as a principle is one of the most widely studied properties in the axiomatic allocation literature. While the implementation of consistency varies depending on the application, the general principle is that a rule should not alter the allocation among a group when other agents are removed from a problem in such a way so that the new reduced problem is similar in nature to the original problem. The principle is applied in Nash bargaining problems \citep{lensberg1987scr}, coalitional games \citep{hart1989pva}, claims problems \citep{aumann1985gta}, and matching problems \citep{sasaki1992consistency}.\footnote{As discussed in \autoref{Sec_Coalitions}, compensation problems can be viewed as a coalitional game with the characteristic function $v(S) = y(S,T,p_S)$ for $S \subset I$. However, our Consistency axiom is different from the consistency axiom employed by \citet{hart1989pva}. The key difference is in how the reduced problem is defined. In a compensation problem, the added structure afforded by the tasks provides a natural way to define the reduced problem. However, this added structure means that it is not possible to define the reduced problem in an equivalent manner using just the characteristic function $v$.} See \citet{thomson2011cai} for a short introduction and \citet{thomson2013car} for a comprehensive treatment of the literature on the consistency principle.

While Weak Consistency is a straightforward application of the consistency principle to this setting, it does not address those compensation problems that have multiple optimal assignments. The difficulty with applying the consistency principle to such rules is that it is not evident which optimal assignment to use when defining the reduced problem. To illustrate this point, suppose $(I,T,p)$ has multiple optimal assignments. Then for $a,\hat{a} \in A^*(I,T,p)$ and $I' \subset I$, the set of tasks $a(I')$ and $\hat{a}(I')$ are potentially different, and thus the problems $(I',a(I'),p_{I'}^{a(I')})$ and $(I',\hat{a}(I'),p_{I'}^{\hat{a}(I')})$ could be completely different. In particular, there is no guarantee that $y(I',a(I'),p_{I'}^{a(I')})$ and $y(I',\hat{a}(I'),p_{I'}^{\hat{a}(I')})$ would be equal. 

Thus any attempt to extend Weak Consistency to compensation problems with multiple optimal assignments must address this issue. The following axiom strengthens Weak Consistency by using all of a problem's optimal assignments to define the reduced problem. 

\begin{axiom}[Consistency]
For every $(I,T,p)$ and $I' \subset I$, we have 
\[
R_{I'} (I,T,p) = \frac{1}{|A^* (I,T,p)|} \sum_{a \in A^* (I,T,p)} R\left(I',a(I'),p_{I'}^{a(I')}\right) \text{.}
\]
\end{axiom}

When $A^* (I,T,p)$ is a singleton, then Consistency reduces to Weak Consistency. When $A^* (I,T,p)$ is not a singleton, then Consistency insists that the compensation $i \in I'$ receives under $(I,T,p)$ is equal to the average of their compensation under all possible reduced problems for $I'$. 

Consistency fills a gap left by Weak Consistency by addressing how to define the reduced problem when there are multiple optimal assignments. However, we note that this is not the only way to address this issue. In \autoref{Sec_App_IC}, we explore this further by considering other compensation rules that satisfy Weak Consistency but not Consistency. However, we also show that these differences are of minor importance since the set of compensation problems with multiple optimal assignments is, in some sense, sparse. Thus all rules that satisfy Weak Consistency agree on compensation for nearly all problems.

\begin{theorem} \label{thm02}
The Individual Contribution rule is the only compensation rule satisfying Efficiency and Consistency.
\end{theorem}
\begin{proof}
It is straightforward to show that the the Individual Contribution rule satisfies Efficiency. Consistency is established in \autoref{A29} in \autoref{Sec_appendix}.

To show the Individual Contribution rule is the only such rule, let $R$ be a rule that satisfies Efficiency and Consistency. Fix the problem $(I,T,p)$ and individual $i \in I$. Since $R$ satisfies Consistency, we have
\[
R_{i} (I,T,p) = \frac{1}{|A^* (I,T,p)|} \sum_{a \in A^* (I,T,p)} R\left(\{ i \} , a(i) , p_{ i }^{a(i)} \right) \text{.}
\]

Fix $a \in A^* (I,T,p)$. Note that $(\{ i \} , a(i) , p_{ i }^{a(i)} )$ is a problem with one individual, and so Efficiency implies $R(\{ i \} , a(i) , p_{ i }^{a(i)} ) = p_{ i }^{a(i)}$. Substituting this into the above equation, we have 
\[
R_{i} (I,T,p) = \frac{1}{|A^* (I,T,p)|} \sum_{a \in A^* (I,T,p)}  p_{i}^{a(i)} \text{,}
\]
which is the Individual Contribution rule.
\end{proof}

The Individual Contribution rule is arguably fair since each worker receives what they personally contribute to the output. However, as discussed in the introduction, this can lead to some arguably unfair outcomes since an individual's contribution is chosen so as to maximize the collective output. Indeed, as will be shown in the next section, the Individual Contribution rule violates a number of normatively appealing axioms.

\section{Other Rules and Axioms} \label{sec_Axioms}

In this section, we introduce a number of other compensation rules, as well as other axioms. We relate each axiom to each compensation rule. Our results are summarized in \autoref*{table02}. Only a selection of these results are justified in the discussion that follows. See \autoref{Sec_appendix} for justifications for other entries in this table.

\begin{table}
\centering
\begin{threeparttable}
  \begin{tabular}{  l  p{.32in}  p{.32in}  p{.32in}  p{.32in}  p{.32in}  p{.32in}  p{.32in}  p{.32in} }
     & $E$ & $SV$ & $IC$ & $P^{Av}$ & $P^{\max}$ & $P^{\Delta}$  & $E^\Delta$ & $Par^f$  \\ \hline \hline 
    Efficiency &  $\oplus$ & $\oplus$ & $\oplus$ & $+$ & $+$ & $+$ & $+$ & $+$   \\ \hline
    Continuity &  $+$ & $+$ & $-$ & $+$ & $+$ & $+$ & $+$ & $\pm$   \\ \hline
    Boundedness &  $-$ & $+$ & $+$ & $-$ & $-$ & $-$ & $-$ & $\pm$   \\ \hline
    Symmetry &  $\oplus$ & $+$ & $+$ & $+$ & $+$ & $+$ & $+$ & $+$  \\ 
    $\pi$-Symmetry &  $+$  & $-$ & $-$ & $+$ & $+$ & $-$ & $-$ & $\pm$ \\ 
    Order Preservation &  $+$  & $+$ & $-$ & $+$ & $+$ & $+$ & $+$ & $\pm$  \\
    Strict Order Preservation &  $-$  & $+$ & $-$ & $+$ & $+$ & $+$ & $+$ & $\pm$  \\    
    Strong Order Preservation &  $-$  & $-$ & $-$ & $+$ & $-$ & $-$ & $-$ & $\pm$  \\
    $\pi$-Order Preservation &  $+$ & $-$  & $-$ & $+$ & $+$ & $-$ & $-$ & $\pm$  \\ \hline
    Group Prod. Mono. &  $\oplus$  & $-$  & $-$ & $-$ & $-$ & $-$ & $-$ & $-$ \\ 
    Individual Prod. Mono. &  $+$ & $+$ & $-$ & $+$ & $+$ & $-$ & $+$ & $\pm$ \\
    Strict Individual Prod. Mono. &  $+$ & $+$ & $-$ & $+$ & $+$ & $-$ & $+$ & $\pm$ \\
    Strong Individual Prod. Mono. &  $-$  & $-$  & $-$ & $+$ & $-$ & $-$ & $-$ & $\pm$ \\ \hline
    Constant Productivity &  $-$ & $+$ & $+$ & $-$ & $-$ & $-$ & $-$ & $\pm$  \\
    Trivialness &  $-$ & $+$ & $+$ & $+$ & $+$ & $+$ & $+$ & $\pm$  \\ \hline
    Balanced Impact &  $-$ & $\oplus$ & $-$ & $-$ & $-$ & $-$ & $-$ & $-$ \\   
    Consistency &   $-$ & $-$ & $\oplus$ & $-$ & $-$ & $-$ & $-$ & $-$  \\   
    Ind. Null Workers &  $-$ & $+$ & $+$ & $+$ & $+$ & $+$ & $-$ & $\pm$ \\ 
    No Harm from Hiring &  $-$ & $-$  & $-$ & $-$ & $-$ & $-$ & $-$ & $-$ \\ 
    Solidarity in Hiring &  $+$ & $+$ & $-$ & $+$ & $+$ & $-$ & $-$ & $+$ \\    \hline
    Ind. Null Tasks &  $+$ & $+$ & $-$ & $+$ & $+$ & $+$ & $+$ & $\pm$ \\ 
    Ind. Unassigned Tasks &  $+$ & $+$ & $+$ & $-$ & $+$ & $+$ & $+$ & $\pm$ \\  \hline 
    Additivity &  $+$ & $-$ & $-$ & $+$ & $-$ & $-$ & $-$ & $\pm$ \\
    Weak Additivity &  $+$ & $-$ & $+$ & $+$ & $-$ & $-$ & $-$  & $\pm$ \\
    Homogeneity &  $+$ & $+$ & $+$ & $+$ & $+$ & $+$ & $+$ & $\pm$   \\

%    \hline \hline
%    Source & \lbrack 1\rbrack & \lbrack 2\rbrack & \lbrack 3\rbrack & \lbrack 4\rbrack & \lbrack 5\rbrack & \lbrack 6\rbrack \\ 
  \end{tabular}
%  \begin{tablenotes}[flushleft]
%  \item [\lbrack 1\rbrack] \autoref{thm01}.
%  \item [\lbrack 2\rbrack] \autoref{thm02}.
%  \item [\lbrack 3\rbrack] \autoref{thm03}.
%  \item [\lbrack 4\rbrack] \autoref{ex02}.
%  \item [\lbrack 5\rbrack] \autoref{ex03}. 
%  \item [\lbrack 6\rbrack] \autoref{ex04}.
%  \item [\lbrack 7\rbrack] \autoref{ex05}.
%  \item [\lbrack 8\rbrack] \autoref{ex08}.
%  \item [\lbrack 9\rbrack] \autoref{prop01}.
%  \item [\lbrack 10\rbrack] \autoref{ex06}.
%  \end{tablenotes}
\caption{\label{table02} \textbf{Summary of rules and axioms.} The symbols $+$ and $-$ indicate the axiom is necessary and not necessary, respectively. For the first three columns, the set of axioms indicated by $\oplus$ are necessary and sufficient for the given compensation rule, as given by the results in \autoref{Sec_Results}. For the last column, entries of $\pm$ indicate that the axiom could be satisfied under suitable restrictions on the parametric function $f$.} 
\end{threeparttable}
\end{table}

\subsection{Other Compensation Rules}

For a fixed problem $(I,T,p)$ and for $i \in I$, define
\begin{align*}
\bar{p}_i &\coloneqq \frac{1}{|T|} \sum_{t \in T} p_i^t \text{,} \\
p_i^{\max} &\coloneqq \max_{t \in T} p_i^t \text{, and} \\
\Delta p_i &\coloneqq y(I,T,p) - y(I \setminus \{ i \} , T, p_{-i}) \text{.}
\end{align*}
Thus $\bar{p}_i$ is individual $i$'s average productivity, $p_i^{\max}$ is their maximum productivity, and $\Delta p_i$ is their marginal contribution. Each of these values can be thought of as a single variable summary of individual $i$'s productivity in the problem $(I,T,p)$. We consider three compensation rules that each assign compensation proportionally using these three measures of productivity.

\begin{axiom}[Proportional to Average Productivity Rule]
For any problem $(I,T,p)$ and for any $i \in I$, the Proportional to Average Productivity rule $P^{Av}$ assigns to $i$ the compensation
\[
P^{Av}_i (I,T,p) \coloneqq \frac{y(I,T,p)}{\sum_{j \in I} \bar{p}_j} \bar{p}_i \text{.}
\]
\end{axiom}

\begin{axiom}[Proportional to Max Productivity Rule]
For any problem $(I,T,p)$ and for any $i \in I$, the Proportional to Max Productivity rule $P^{\max}$ assigns to $i$ the compensation
\[
P^{\max}_i (I,T,p) \coloneqq \frac{y(I,T,p)}{\sum_{j \in I} p_j^{\max}} p_i^{\max} \text{.}
\]
\end{axiom}

\begin{axiom}[Proportional to Marginal Contribution Rule]
For any problem $(I,T,p)$ and for any $i \in I$, the Proportional to Marginal Contribution rule $P^{\Delta}$ assigns to $i$ the compensation
\[
P^{\Delta}_i (I,T,p) \coloneqq \frac{y(I,T,p)}{\sum_{j \in I} \Delta p_j} \Delta p_i \text{.}
\]
\end{axiom}

Note that in general $\frac{y(I,T,p)}{\sum_{j \in I} \Delta p_j} \geq 1$ since $\sum_{j \in I} \Delta p_j \leq y(I,T,p)$. Thus one way to think of $P^{\Delta}$ is that each individual receives their respective marginal contribution, and then the surplus is distributed proportionally to their marginal contribution. The following rule instead distributes that surplus equally.

\begin{axiom}[Marginal Egalitarian Rule]
For any problem $(I,T,p)$ and for any $i \in I$, the Marginal Egalitarian rule $E^\Delta$ assigns to $i$ the compensation
\[
E^\Delta_i (I,T,p) \coloneqq \Delta p_i + \frac{1}{|I|} \left( y(I,T,p) - \sum_{j \in I} \Delta p_j \right) \text{.}
\]
\end{axiom}

Finally, we introduce a family of parametric compensation rules similar to the family introduced by \citet{young1987daa} for the conflicting claims problem.

\begin{axiom}[Parametric Rule]
Let $f: \cup_{T \in \mathcal{N}} \mathbb{R}^T \times \mathbb{R}_+ \rightarrow \mathbb{R}_+$ satisfy: (i) $f(p_0, \lambda)$ is continuous in $\lambda$; (ii) $f(p_0, \lambda)$ is weakly monotone in $\lambda$; (iii) $f(p_0 , 0) = 0$ for every $p_0 \in \cup_{T \in \mathcal{N}} \mathbb{R}^T$; and (iv) $\lim_{\lambda \rightarrow \infty} f(p_0,\lambda) = \infty$ for every $p_0 \in \cup_{T \in \mathcal{N}} \mathbb{R}^T$ that is not null (i.e. $p_0 \neq (0, \ldots , 0)$). For any problem $(I,T,p)$ and for any $i \in I$, $Par^f$ assigns to $i$ the compensation
\[
Par^f_i (I,T,p) \coloneqq f(p_i, \lambda) \text{,}
\]
where $\lambda$ is chosen so that $\sum_{j \in I} f(p_j, \lambda) = y(I,T,p)$.\footnote{Because $f$ is continuous, weakly monotone, and unbounded from above with respect to the second argument, such a $\lambda$ always exists and leads to a unique solution.}
\end{axiom}

Of the rules we have considered above, $E$, $P^{Av}$, and $P^{\max}$ are members of the Parametric family, while the others are not. Obviously, every Parametric rule will satisfy Efficiency and Symmetry. However, without additional restrictions on the parametric function $f$, $Par^f$ will not satisfy any of the axioms we consider below, with one exception.

\subsection{Other Axioms}

\subsubsection{Continuity}

We consider a continuity axiom. For this axiom, we endow $\mathbb{R}_{+}^{I \times T}$ and $\mathbb{R}^I_+$ with the Euclidean topology.

\begin{axiom}[Continuity]
For every $(I,T,p)$, if $\{p(n)\}$ is a sequence of productivity matrices in $\mathbb{R}_{+}^{I \times T}$ satisfying $p(n) \rightarrow p$, then $R(I,T,p(n)) \rightarrow R(I,T,p)$.
\end{axiom}

Aside from being analytically useful, Continuity has normative appeal. It would seem grossly unfair if an individual's compensation jumped dramatically due to a small change in someone's productivity.

A Parametric rule $Par^f$ is not guaranteed to satisfy Continuity since $f$ is not necessarily continuous in the productivity profile. All of the other compensation rules satisfy Continuity, except for $IC$.\footnote{\autoref*{ex02} can be modified easily to demonstrate that the Weak Individual Contribution rules (introduced in \autoref{Sec_App_IC}) are generally incompatible with Continuity. Let $\hat{p}(n)$ be a sequence of problems where $\hat{p}_{1}(n) = (6 , 4)$ and $\hat{p}_{2}(n) = (3 - \frac{1}{n} , 1)$. Then $\hat{p}(n) \rightarrow p$ also. But if $R$ is a Weak Individual Contribution rule, then we must have $R(I,T,p(n)) = (4,3)$ and $R(I,T,\hat{p}(n)) = (6,1)$ for every $n$. Thus Continuity is violated no matter what $R(I,T,p)$ is. This also demonstrates, in conjunction with \autoref{prop03}, that there is no compensation rule that satisfies Efficiency, Weak Consistency, and Continuity.}

\begin{example} \label{ex02}
\textbf{$IC$ violates Continuity.} Let $I = T = \{1,2\}$. Let $p(n)$ be a sequence of problems where $p_{1}(n) = (6- \frac{1}{n} , 4)$ and $p_{2}(n) = (3 , 1)$. Then $p(n) \rightarrow p$ where $p_1 = (6,4)$ and $p_2 = (3,1)$. But $IC(I,T,p(n)) = (4,3)$ for every $n$, while $IC (I,T,p) = (5,2)$.
\end{example}

\subsubsection{Boundedness}

The next axiom places bounds on how much compensation a worker can receive. For this axiom, we define $p_i^{\min} \coloneqq \min_{t \in T} p_i^t$.

\begin{axiom}[Boundedness]
For every $(I,T,p)$ and for every $i \in I$, we have $p_i^{\min} \leq R_i (I,T,p) \leq p_i^{\max}$.
\end{axiom}

Since $p_i^{\min}$ is the absolute lowest amount $i$ can produce, there is no disputing that they contribute at least $p_i^{\min}$. Similarly, $i$ cannot contribute more than $p_i^{\max}$. Boundedness thus restricts $i$'s compensation to be between these amounts. Among the rules that we consider, only $IC$ and $SV$ satisfy Boundedness.  A Parametric rule $Par^f$ would satisfy Boundedness if there existed $0<a<b$ such that $f(p_0,a) = p_0^{\min}$ and $f(p_0,b) = p_0^{\max}$ for every $p_0 \in \mathbb{R}_+^T$.

\subsubsection{Symmetry and Order Preservation}

Next we consider some axioms related to Symmetry. Imagine a scenario in which two individuals do not have exactly identical productivity profiles task for task, but when their respective productivity profiles are ordered from most to least productive, then they do look identical. One of these individuals may be more useful to the firm depending on the productivity profiles of the other workers, but this could potentially be either of these individuals. The next axiom states that such individuals should receive the same compensation.

\begin{axiom}[$\pi$-Symmetry]
For every $(I,T,p)$ and for every $i,j \in I$, if there exists $\pi$ a permutation\footnote{For any set $X$, a permutation of $X$ is a bijective function $\pi : X \rightarrow X$.} of $T$ such that $p_{i}^{t} = p_{j}^{\pi(t)}$ for every $t \in T$, then $R_i(I,T,p) = R_j (I,T,p)$.
\end{axiom}

Obviously, $\pi$-Symmetry implies Symmetry. Of the rules we consider, $E$, $P^{Av}$, and $P^{\max}$ satisfy $\pi$-Symmetry. The other rules do not.

The next axiom says that the ordering of compensation should coincide with the ordering of productivity profiles.

\begin{axiom}[Order Preservation]
For every $(I,T,p)$ and for every $i,j \in I$, if $p_i \geqq p_j$, then $R_i(I,T,p) \geq R_j (I,T,p)$.
\end{axiom}

Order Preservation implies Symmetry. All of our compensation rules satisfy Order Preservation, except for $IC$ and the Parametric rules. A Parametric rule $Par^f$ would satisfy Order Preservation if $f$ was monotone in the productivity profile. The following example shows a violation of Order Preservation by $IC$.\footnote{\autoref*{ex04} also applies to the Weak Individual Contribution rules introduced in \autoref{Sec_App_IC}, and thus those rules do not satisfy Order Preservation.}

\begin{example} \label{ex04}
\textbf{$IC$ violates Order Preservation.} Let $I = T = \{1,2\}$. Let $p_1 = (4,1)$ and $p_2 = (5,3)$. Then $p_2 > p_1$ and $IC(I, T, p) = (4,3)$.
\end{example}

We consider two stronger versions of Order Preservation that each specify when one individual's compensation is strictly more than another's.

\begin{axiom}[Strict Order Preservation]
For every $(I,T,p)$ and for every $i,j \in I$, if $p_i > p_j$, then $R_i(I,T,p) > R_j (I,T,p)$.
\end{axiom}

\begin{axiom}[Strong Order Preservation]
For every $(I,T,p)$ and for every $i,j \in I$, if $p_i \geq p_j$, then $R_i(I,T,p) > R_j (I,T,p)$.
\end{axiom}

Strong Order Preservation implies Strict Order Preservation. $E$, $IC$, and the Parametric rules satisfy neither axiom. Among our rules, only $P^{Av}$ satisfies both axioms. The remaining rules satisfy Strict but not Strong Order Preservation.

Similar to $\pi$-Symmetry, the next axiom imposes the Order Preservation condition on permutations of the productivity profile.

\begin{axiom}[$\pi$-Order Preservation]
For every $(I,T,p)$ and for every $i,j \in I$, if there exists $\pi$ a permutation over $T$ such that $p_{i}^{t} \geq p_{j}^{\pi(t)}$ for every $t \in T$, then $R_i(I,T,p) \geq R_j (I,T,p)$.
\end{axiom}

$\pi$-Order Preservation implies both Order Preservation and $\pi$-Symmetry. $E$, $P^{Av}$, and $P^{\max}$ all satisfy $\pi$-Order Preservation, while the other rules do not.

\subsubsection{Individual Productivity Monotonicity} \label{Sec_Mono}

The next axiom weakens Group Productivity Monotonicity so that only the worker whose productivity increases is guaranteed an increase in compensation.

\begin{axiom}[Individual Productivity Monotonicity]
For every $(I,T,p)$ and for every $i \in I$, if $p_i \geq \hat{p}_i $, then $R_i(I,T,p) \geq R_i(I,T,(p_{-i},\hat{p}_i))$.
\end{axiom}

All of our compensation rules satisfy Individual Productivity Monotonicity, except for $IC$, $P^{\Delta}$, and the Parametric rules. A Parametric rule $Par^f$ would satisfy Individual Productivity Monotonicity if $f$ was monotone in the productivity profile. The following example shows a violation of Individual Productivity Monotonicity by $IC$.\footnote{\autoref*{ex05} also applies to the Weak Individual Contribution rules introduced in \autoref{Sec_App_IC}, and thus those rules do not satisfy Individual Productivity Monotonicity.}

\begin{example} \label{ex05}
\textbf{$IC$ violates Individual Productivity Monotonicity.} Continuing \autoref{ex04}, let $\hat{p}_2 = (4,0)$. Then  $p_2 > \hat{p}_2$ and $IC_2(I,T,(p_{1},\hat{p}_2)) = 4 > 3 = IC_2(I, T, p)$.
\end{example}

We consider two stronger versions of Individual Productivity Monotonicity that each specify when an individual's compensation should strictly increase due to an increase in productivity.

\begin{axiom}[Strict Individual Productivity Monotonicity]
For every $(I,T,p)$ and for every $i \in I$, if $p_i > \hat{p}_i $, then $R_i(I,T,p) > R_i(I,T,(p_{-i},\hat{p}_i))$.
\end{axiom}

\begin{axiom}[Strong Individual Productivity Monotonicity]
For every $(I,T,p)$ and for every $i \in I$, if $p_i \geq \hat{p}_i $, then $R_i(I,T,p) > R_i(I,T,(p_{-i},\hat{p}_i))$.
\end{axiom}

Strong Individual Productivity Monotonicity implies Strict Individual Productivity Monotonicity. $IC$, $P^{\Delta}$, and the Parametric rules satisfy neither axiom. Among the rules we consider, only $P^{Av}$ satisfies both axioms. The remaining rules satisfy Strict but not Strong Individual Productivity Monotonicity.

\subsubsection{Constant Productivity}

The axioms in this section consider compensation problems in which an individual has constant productivity. For problem $(I,T,p)$, we say $i \in I$ has \emph{constant productivity $\alpha \in \mathbb{R}_+$} if $p_i^t = \alpha$ for every $t \in T$. When there is such an individual in a compensation problem, then there is no debate about what she adds to the firm's output. Our first axiom states that the individual's constant productivity should be her compensation.

\begin{axiom}[Constant Productivity]
For every $(I,T,p)$, if $i \in I$ has constant productivity $\alpha$, then $R_i(I,T,p) = \alpha$.
\end{axiom}

Boundedness implies Constant Productivity. Of our rules, only $IC$ and $SV$ satisfy Constant Productivity. 

The next axiom is weaker than Constant Productivity. We say $(I,T,p)$ is a \emph{trivial problem} if every individual has constant productivity. That is, for every $i \in I$, there exists $\alpha_i \in \mathbb{R}_+$ such that $i$ has constant productivity $\alpha_i$. In what follows, we will denote a trivial problem as $(I,T,(\alpha_i )_{i \in I})$. For a trivial problem, the assignment of tasks is irrelevant and the overall contribution of each worker is unambiguous.

\begin{axiom}[Trivialness]
For every trivial problem $(I,T,(\alpha_i )_{i \in I})$, we have $R_i(I,T,(\alpha_i )_{i \in I}) = \alpha_i$ for every $i \in I$.
\end{axiom}

Obviously Constant Productivity implies Trivialness. Of our compensation rules, only $E$ and the Parametric rules do not satisfy Trivialness.

\subsubsection{One Worker's Effect on the Group} \label{Sec_NHfH}

Both Consistency and Balanced Impact concern how one worker's presence affects another worker's compensation. We consider a few more axioms along these lines.

For problem $(I,T,p)$, we say $i \in I$ is a \emph{null worker} if $p_i^t = 0$ for every $t \in T$. Constant Productivity says that a null worker should get zero compensation. The next axiom says that a null worker should have no effect on the other workers.

\begin{axiom}[Independence of Null Workers]
For every $(I,T,p)$, if $i \in I$ is a null worker, then $R_{-i}(I,T,p) = R(I \setminus \{ i \},T,p_{-i})$.
\end{axiom}

Independence of Null Workers could be thought of as a very weak form of Consistency since it states that the compensation of a subgroup of workers should not change when one (very specific) worker is removed. Among our compensation rules, $E$, $E^\Delta$, and the Parametric rules do not satisfy Independence of Null Workers, while the other rules do. A Parametric rule $Par^f$ would satisfy Independence of Null Workers if $f$ satisfied $f((0 , \ldots , 0),\lambda) = 0$ for every $\lambda$.

The next axiom states that workers should not be made worse off when the firm hires new workers.

\begin{axiom}[No Harm from Hiring]
For every $(I,T,p)$ and for every $I' \subset I$, we have $R_{I'}(I,T,p) \geqq R (I',T,p_{I'})$.
\end{axiom}

While No Harm from Hiring has normative appeal, unfortunately it is incompatible with Efficiency. This is because the output function $y$ is generally submodular with respect to the set of workers. That is for any problem $(I,T,p)$, and for any bipartition $\{J,K\}$ of $I$,\footnote{A bipartition of a set $A$ are subsets $X$ and $Y$ such that $X \cup Y = A$ and $X \cap Y = \emptyset$.} we have
\[
y(J,T,p_J) + y(K,T,p_K) \geq y(I,T,p) \text{.}
\] 
Any problem in which the above inequality is strict would lead to a contradiction between Efficiency and No Harm from Hiring, as the following example illustrates.

\begin{example}
\textbf{Efficiency and No Harm from Hiring are incompatible.} Return to the compensation problem from \autoref{ex04}. Note that if $R$ satisfies Efficiency, then $R(\{1\},T,p_1) = 4$ and $R(\{2\},T,p_2) = 5$. Thus No Harm from Hiring implies $R_1(I,T,p) \geq 4$ and $R_2(I,T,p) \geq 5$. But Efficiency implies $R_1(I,T,p) + R_2(I,T,p) = 7$,
which is a contradiction.
\end{example}

Thus we have the following result.

\begin{proposition} \label{prop01}
No compensation rule satisfies Efficiency and No Harm from Hiring.
\end{proposition}

A weaker condition requires that when someone is hired, the original workers are affected in the same direction.\footnote{Solidarity in Hiring is similar to the Population-and-Resource Monotonicity axiom introduced by \citet{chun1999eoa} for the conflicting claims problem.}

\begin{axiom}[Solidarity in Hiring]
For every $(I,T,p)$ and for every $I' \subset I$, we have either $R_{I'}(I,T,p) \leqq R (I',T,p_{I'})$ or $R_{I'}(I,T,p) \geqq R (I',T,p_{I'})$.
\end{axiom}

$IC$, $P^{\Delta}$, and $E^\Delta$ do not satisfy Solidarity in Hiring. The following example demonstrates this for $IC$.\footnote{\autoref*{ex06} also applies to the Weak Individual Contribution rules introduced in \autoref{Sec_App_IC}, and thus those rules do not satisfy Solidarity in Hiring.}

\begin{example} \label{ex06}
\textbf{$IC$ violates Solidarity in Hiring.} Let $I = T = \{1,2,3\}$ and $I' = \{1,2\}$. Let $p_1 = (4,1,3)$, $p_2 = (4,2,1)$, and $p_3 = (1,1,4)$. Then $IC(I', T, p_{-3}) = (3,4)$, while $IC_{I'}(I, T, p) = (4,2)$.
\end{example}

All of our other rules satisfy Solidarity in Hiring. In particular, this is the only axiom that we consider --- besides Efficiency and Symmetry --- that the Parametric rules generally satisfy.

\subsubsection{Changing Tasks} 

Our next set of axioms concern changes in the set of tasks. For problem $(I,T,p)$, we say $t \in T$ is a \emph{null task} if $p_{i}^{t} =0$ for every $i \in I$.

\begin{axiom}[Independence of Null Tasks]
For every $(I,T,p)$ where $|T| \geq |I|+1$, if $t \in T$ is a null task, then $R(I,T,p) = R(I,T \setminus \{t\},p^{-t})$.
\end{axiom}

All of our compensation rules satisfy Independence of Null Tasks, except for $IC$ and the Parametric rules. A Parametric rule $Par^f$ would satisfy Independence of Null Tasks if $f$ satisfied the following condition: Whenever $\hat{T} \subset T$, $p_0 \in \mathbb{R}_+^T$, $\hat{p}_0 \in \mathbb{R}_+^{\hat{T}}$, $p_0^t = \hat{p}_0^t$ for every $t \in \hat{T}$, and $p_0^t = 0$ for every $t \not\in \hat{T}$, we have $f(p_0,\lambda) = f(\hat{p}_0,\lambda)$ for every $\lambda$.\footnote{One example of such a parametric function is $f(p_0 , \lambda) = \lambda \sum_{t \in T} p_0^t$.} The following example shows a violation of Independence of Null Tasks by $IC$.

\begin{example} \label{ex07} 
\textbf{$IC$ violates Independence of Null Tasks.} Let $I = \{1,2\}$ and $ T = \{1,2,3\}$. Let $p_1 = (2,1,0)$ and $p_2 = (1,0,0)$. Then $3 \in T$ is a null task, yet $IC(I, T, p) = (\frac{5}{3}, \frac{1}{3})$ and $IC(I, T \setminus \{3\}, p^{-3}) = (\frac{3}{2}, \frac{1}{2})$.
\end{example}

For problem $(I,T,p)$, let $T^* (I,T,p)$ denote the set of tasks that will be performed under some optimal assignment. I.e.
\[
T^* (I,T,p) \coloneqq \{ a(i) \in T : a \in A^* (I,T,p) \text{ and } i \in I  \} \text{.}
\]

\begin{axiom}[Independence of Unassigned Tasks]
For every $(I,T,p)$, for every $T'$ satisfying $T^* (I,T,p) \subset T' \subset T$, we have $R(I,T,p) = R(I,T',p^{T'})$.
\end{axiom}

Of our compensation rules, only $P^{Av}$ and the Parametric rules do not satisfy Independence of Unassigned Tasks. A Parametric rule $Par^f$ would satisfy Independence of Unassigned Tasks if $f$ satisfied $f(p_0,\lambda) = f(p_0^{\max},\lambda)$ for every $\lambda$.

\subsubsection{Additivity} \label{Sec_Add}

Our final set of axioms impose additivity conditions with respect to the productivity matrix. 

\begin{axiom}[Additivity]
For every $I,T \in \mathcal{N}$ where $|I| \leq |T|$, for every $p,\hat{p} \in \mathbb{R}_{+}^{I \times T}$ satisfying $A^* (I,T,p) \cap A^* (I,T,\hat{p}) \neq \emptyset$, we have $R(I,T,p+\hat{p}) = R(I,T,p) + R(I,T,\hat{p})$.
\end{axiom}

The condition $A^* (I,T,p) \cap A^* (I,T,\hat{p}) \neq \emptyset$ serves two purposes. First, when the two compensation problems share an optimal assignment of tasks, then it makes the additivity requirement more natural since each worker could do the same task across the different problems. Second, without this condition, the additivity requirement would conflict with Efficiency since we could have $y(I,T,p) + y(I,T,\hat{p}) > y(I,T,p+\hat{p})$.

Of our main compensation rules, only $E$ and $P^{Av}$ satisfy Additivity. A Parametric rule $Par^f$ would satisfy Additivity if $f$ was additive with respect to the productivity profile. I.e. for every $T$, for every $p_0,\hat{p}_0 \in \mathbb{R}_+^T$, and for every $\lambda$, we have $f(p_0 + \hat{p}_0, \lambda) = f(p_0 , \lambda) + f(\hat{p}_0, \lambda)$. The following example demonstrates the negative result for $IC$.

\begin{example} \label{ex08}
\textbf{$IC$ violates Additivity.} Let $I = T = \{1,2\}$. Let $p_1 = p_2 =(2,1)$, $\hat{p}_1 = (1,0)$, and $\hat{p}_2 = (0,1)$. Note that the assignment $a(i)=i$ satisfies $a \in A^* (I,T,p) \cap A^* (I,T,\hat{p})$. Note also that $IC(I, T, p)  = (\frac{3}{2} , \frac{3}{2})$ and $IC(I, T, \hat{p}) = (1,1)$. However, $IC(I, T, p + \hat{p}) = (3,2)$.
\end{example}

The following weakening imposes the additivity requirement only when the sets of optimal assignments exactly coincide.

\begin{axiom}[Weak Additivity]
For every $I,T \in \mathcal{N}$ where $|I| \leq |T|$, for every $p,\hat{p} \in \mathbb{R}_{+}^{I \times T}$ satisfying $A^* (I,T,p) = A^* (I,T,\hat{p})$, we have $R(I,T,p+\hat{p}) = R(I,T,p) + R(I,T,\hat{p})$.
\end{axiom}

Obviously Additivity implies Weak Additivity. In addition to $E$ and $P^{Av}$, $IC$ also satisfies Weak Additivity, while our other rules do not. The following example demonstrates the negative result for $SV$.

\begin{example} \label{ex09}
\textbf{$SV$ violates Weak Additivity.} Let $I = T = \{1,2\}$. Let $p_1 =(3,0)$, $p_2 =(2,1)$, $\hat{p}_1 = (1,0)$, and $\hat{p}_2 = (0,1)$. Note that the assignment $a(i)=i$ is the unique optimal assignment for both $(I,T,p)$ and $(I,T,\hat{p})$. Note also that $SV(I, T, p) = (\frac{5}{2} , \frac{3}{2})$ and $SV(I, T, \hat{p}) = (1,1)$. However, $SV(I, T, p + \hat{p}) = (4,2)$.
\end{example}

This result is in contrast to Shapley's original characterization which employed an additivity requirement in the coalitional game setting \citep{shapley1953avf}.

Our final axiom states that if productivity is scaled up by some factor, then compensation should scale up by the same factor.

\begin{axiom}[Homogeneity]
For every $(I,T,p)$ and for every $\alpha >0$, we have $R(I,T,\alpha p) = \alpha R(I,T,p)$.
\end{axiom}

It is not difficult to show that Continuity and Weak Additivity imply Homogeneity. Parametric rules do not generally satisfy Homogeneity. A Parametric rule $Par^f$ would satisfy Homogeneity if $f$ was homogeneous of degree one with respect to the productivity profile.  All of our other compensation rules satisfy Homogeneity.

\section{Discussion}

Many of the simplifying assumptions in our model make the problem more tractable, but also point to paths of future research. For example, our production process rules out synergies that may exist between workers. Indeed, as pointed out above, the output function $y$ is generally submodular with respect to the group of workers. While allowing for synergies would be an interesting extension of the problem, it would also present some challenges in formulating axioms. For example, the normative appeal of Consistency rests on the fact that much of the compensation problem remains constant when the group of workers shrinks. However, when an individual's productivity profile is allowed to change when the group of workers changes, then the consistency property becomes less defensible.

Another important assumption is that workers have no control over their own productivity. However, if workers could control their output levels through costly effort, then the compensation rule could affect the amount produced. For example, a hospital may find that emergency room wait times change drastically depending on if they pay their doctors a fixed amount per shift (i.e. the Egalitarian rule) or if they pay according to the number of patients a doctor sees (i.e. the Individual Contribution rule).

Finally, our analysis says nothing about how compensation rules are determined in a competitive environment in which workers are free to choose which firm they work for. An important question, therefore, is what compensation rules survive in equilibrium when firms compete for workers.

\clearpage

%%%%%
%%Appendix
%%%%%

\appendix

\section*{Appendix}

\section{Proofs for Selected Results}  \label{Sec_appendix}

\begin{table}[!h]
\centering
\begin{threeparttable}
  \begin{tabular}{  l p{.35in} p{.35in} p{.35in} p{.35in} p{.35in} p{.35in} p{.35in} p{.35in} }
     & $E$ & $SV$ & $IC$ & $P^{Av}$ & $P^{\max}$ & $P^{\Delta}$  & $E^\Delta$ & $Par^f$  \\ \hline \hline 
    Efficiency &  $+$ & $+$ & $+$ & $+$ & $+$ & $+$ & $+$ & $+$    \\ \hline
    Continuity &  $+$ & $+$ & $-$\tnote{\lbrack \ref{A02}\rbrack} & $+$ & $+$ & $+$ & $+$ & $\pm$   \\ \hline
    Boundedness &  $-$ & $+$\tnote{\lbrack \ref{A02a}\rbrack} & $+$ & $-$\tnote{\lbrack \ref{A02b}\rbrack} & $-$\tnote{\lbrack \ref{A02b}\rbrack} & $-$\tnote{\lbrack \ref{A02b}\rbrack} & $-$\tnote{\lbrack \ref{A02b}\rbrack} & $\pm$   \\ \hline
    Symmetry &  $+$ & $+$ & $+$ & $+$ & $+$ & $+$ & $+$ & $+$  \\ 
    $\pi$-Symmetry &  $+$  & $-$\tnote{\lbrack \ref{A04}\rbrack} & $-$\tnote{\lbrack \ref{A04}\rbrack} & $+$ & $+$ & $-$\tnote{\lbrack \ref{A04}\rbrack} & $-$\tnote{\lbrack \ref{A04}\rbrack} & $\pm$  \\ 
    Order Preservation &  $+$  & $+$\tnote{\lbrack \ref{A07}\rbrack} & $-$\tnote{\lbrack \ref{A06}\rbrack} & $+$ & $+$ & $+$ & $+$ & $\pm$ \\
    Strict Order Preservation &  $-$  & $+$\tnote{\lbrack \ref{A12}\rbrack} & $-$\tnote{\lbrack \ref{A06}\rbrack} & $+$ & $+$ & $+$ & $+$ & $\pm$  \\    
    Strong Order Preservation &  $-$  & $-$\tnote{\lbrack \ref{A13}\rbrack} & $-$\tnote{\lbrack \ref{A06}\rbrack} & $+$ & $-$\tnote{\lbrack \ref{A13}\rbrack} & $-$\tnote{\lbrack \ref{A13}\rbrack} & $-$\tnote{\lbrack \ref{A13}\rbrack} & $\pm$  \\
    $\pi$-Order Preservation &  $+$ & $-$\tnote{\lbrack \ref{A04}\rbrack}  & $-$\tnote{\lbrack \ref{A04}\rbrack} & $+$ & $+$ & $-$\tnote{\lbrack \ref{A04}\rbrack} & $-$\tnote{\lbrack \ref{A04}\rbrack} & $\pm$  \\ \hline
    Group Prod. Mono. &  $+$  & $-$\tnote{\lbrack \ref{A15}\rbrack}  & $-$\tnote{\lbrack \ref{A15}\rbrack} & $-$\tnote{\lbrack \ref{A15}\rbrack} & $-$\tnote{\lbrack \ref{A15}\rbrack} & $-$\tnote{\lbrack \ref{A15}\rbrack} & $-$\tnote{\lbrack \ref{A15}\rbrack} & $-$\tnote{\lbrack \ref{A15}\rbrack} \\ 
    Individual Prod. Mono. &  $+$  & $+$\tnote{\lbrack \ref{A18}\rbrack} & $-$\tnote{\lbrack \ref{A17}\rbrack} & $+$ & $+$ & $-$\tnote{\lbrack \ref{A21A}\rbrack} & $+$\tnote{\lbrack \ref{A21}\rbrack} & $\pm$ \\
    Strict Individual Prod. Mono. &  $+$ & $+$\tnote{\lbrack \ref{A22}\rbrack} & $-$\tnote{\lbrack \ref{A17}\rbrack} & $+$ & $+$ & $-$\tnote{\lbrack \ref{A21A}\rbrack} & $+$\tnote{\lbrack \ref{A22a}\rbrack} & $\pm$ \\
    Strong Individual Prod. Mono. &  $-$ & $-$\tnote{\lbrack \ref{A23}\rbrack} & $-$\tnote{\lbrack \ref{A17}\rbrack} & $+$ & $-$\tnote{\lbrack \ref{A23}\rbrack}  & $-$\tnote{\lbrack \ref{A23}\rbrack}  & $-$\tnote{\lbrack \ref{A23}\rbrack} & $\pm$ \\ \hline
    Constant Productivity &  $-$ & $+$\tnote{\lbrack \ref{A02a}\rbrack} & $+$ & $-$\tnote{\lbrack \ref{A02b}\rbrack} & $-$\tnote{\lbrack \ref{A02b}\rbrack} & $-$\tnote{\lbrack \ref{A02b}\rbrack} & $-$\tnote{\lbrack \ref{A02b}\rbrack} & $\pm$ \\
    Trivialness &  $-$ & $+$\tnote{\lbrack \ref{A02a}\rbrack} & $+$ & $+$ & $+$ & $+$ & $+$ & $\pm$ \\ \hline
    Balanced Impact & $-$\tnote{\lbrack \ref{A30}\rbrack} & $+$ & $-$\tnote{\lbrack \ref{A30}\rbrack} & $-$\tnote{\lbrack \ref{A30}\rbrack} & $-$\tnote{\lbrack \ref{A30}\rbrack} & $-$\tnote{\lbrack \ref{A30}\rbrack} & $-$\tnote{\lbrack \ref{A30}\rbrack} & $-$\tnote{\lbrack \ref{A30}\rbrack} \\    
    Consistency & $-$\tnote{\lbrack \ref{A28}\rbrack} & $-$\tnote{\lbrack \ref{A28}\rbrack} & $+$\tnote{\lbrack \ref{A29}\rbrack} & $-$\tnote{\lbrack \ref{A28}\rbrack} & $-$\tnote{\lbrack \ref{A28}\rbrack} & $-$\tnote{\lbrack \ref{A28}\rbrack} & $-$\tnote{\lbrack \ref{A28}\rbrack} & $-$\tnote{\lbrack \ref{A28}\rbrack} \\  
    Ind. Null Workers &  $-$ & $+$\tnote{\lbrack \ref{A34}\rbrack} & $+$\tnote{\lbrack \ref{A33}\rbrack} & $+$ & $+$ & $+$ & $-$ & $\pm$ \\ 
    No Harm from Hiring &  $-$\tnote{\lbrack \ref{A36}\rbrack} & $-$\tnote{\lbrack \ref{A36}\rbrack}  & $-$\tnote{\lbrack \ref{A36}\rbrack} & $-$\tnote{\lbrack \ref{A36}\rbrack} & $-$\tnote{\lbrack \ref{A36}\rbrack} & $-$\tnote{\lbrack \ref{A36}\rbrack} & $-$\tnote{\lbrack \ref{A36}\rbrack} & $-$\tnote{\lbrack \ref{A36}\rbrack} \\ 
    Solidarity in Hiring & $+$\tnote{\lbrack \ref{A39}\rbrack} & $+$\tnote{\lbrack \ref{A38}\rbrack} & $-$\tnote{\lbrack \ref{A37}\rbrack} & $+$\tnote{\lbrack \ref{A39}\rbrack} & $+$\tnote{\lbrack \ref{A39}\rbrack} & $-$\tnote{\lbrack \ref{A40}\rbrack} & $-$\tnote{\lbrack \ref{A40}\rbrack} & $+$\tnote{\lbrack \ref{A39}\rbrack}  \\    \hline
    Ind. Null Tasks &  $+$ & $+$ & $-$\tnote{\lbrack \ref{A42}\rbrack} & $+$ & $+$ & $+$ & $+$ & $\pm$ \\ 
    Ind. Unassigned Tasks & $+$ & $+$ & $+$\tnote{\lbrack \ref{A47}\rbrack} & $-$\tnote{\lbrack \ref{A49}\rbrack} & $+$ & $+$ & $+$ & $\pm$ \\  \hline 
    Additivity & $+$\tnote{\lbrack \ref{A51}\rbrack} & $-$\tnote{\lbrack \ref{A53}\rbrack} & $-$\tnote{\lbrack \ref{A52}\rbrack} & $+$ & $-$\tnote{\lbrack \ref{A55}\rbrack} & $-$\tnote{\lbrack \ref{A55}\rbrack} & $-$\tnote{\lbrack \ref{A55}\rbrack} & $\pm$ \\
    Weak Additivity & $+$\tnote{\lbrack \ref{A51}\rbrack} & $-$\tnote{\lbrack \ref{A53}\rbrack} & $+$\tnote{\lbrack \ref{A56}\rbrack} & $+$ & $-$\tnote{\lbrack \ref{A55}\rbrack} & $-$\tnote{\lbrack \ref{A55}\rbrack} & $-$\tnote{\lbrack \ref{A55}\rbrack}  & $\pm$ \\
    Homogeneity &  $+$ & $+$ & $+$ & $+$ & $+$ & $+$ & $+$  & $\pm$  \\

%    \hline \hline
%    Source & \lbrack 1\rbrack & \lbrack 2\rbrack & \lbrack 3\rbrack & \lbrack 4\rbrack & \lbrack 5\rbrack & \lbrack 6\rbrack \\ 
  \end{tabular}
%  \begin{tablenotes}[flushleft]
%  \item [\lbrack 1\rbrack] \autoref{thm01}.
%  \item [\lbrack 2\rbrack] \autoref{thm02}.
%  \item [\lbrack 3\rbrack] \autoref{thm03}.
%  \item [\lbrack 4\rbrack] \autoref{ex02}.
%  \item [\lbrack 5\rbrack] \autoref{ex03}. 
%  \item [\lbrack 6\rbrack] \autoref{ex04}.
%  \item [\lbrack 7\rbrack] \autoref{ex05}.
%  \item [\lbrack 8\rbrack] \autoref{ex08}.
%  \item [\lbrack 9\rbrack] \autoref{prop01}.
%  \item [\lbrack 10\rbrack] \autoref{ex06}.
%  \end{tablenotes}
\caption{\label{table03} \textbf{Reproduction of \autoref{table02} with references.} Justification for a result is found by matching the reference number to the numbered item in the accompanying list.}
\end{threeparttable}
\end{table}

\autoref{table02} is reproduced in \autoref*{table03} with references. The number next to a result references a justification for that result in the numbered list below. Many results are straightforward and thus left as an exercise to the reader.

\begin{enumerate}

\item \label{A02} This was shown in \autoref{ex02}.

\item \label{A02a} The result follows easily from the following lemma.
\begin{lemma} \label{lemma02}
For every $(I,T,p)$, $p_i^{\min} \leq \Delta p_i \leq p_i^{\max}$.
\end{lemma}
\begin{proof}
Fix the problem $(I,T,p)$. Obviously $\Delta p_i \leq p_i^{\max}$ since $p_i^t \leq p_i^{\max}$ for every $t \in T$.

Moving from the problem $(I \setminus \{i\},T,p_{-i})$ to the problem $(I,T,p)$,
we can always assign $i$ a task $t \in T \setminus a(I \setminus \{i\})$ for some $a \in A^* (I \setminus \{i\},T,p_{-i})$. This would increase output to $y(I \setminus \{i\},T,p_{-i}) + p_i^t$. Hence,
\[
y(I,T,p) \geq y(I \setminus \{i\},T,p_{-i}) +  p_i^t
\]
which implies $\Delta p_i \geq p_i^t$. Finally, since $p_i^t \geq p_i^{\min}$, we get the result.
\end{proof}

\item \label{A02b} The negative result is demonstrated by the following example. Let $I = T = \{1,2,3\}$. Let $p_1 = (1,1,1)$, $p_2 = (2,1,0)$, and $p_3 = (2,0,0)$. Then $P^{Av}_1 (I,T,p) = \frac{3}{2}$, $P^{\max}_1 (I,T,p) = \frac{4}{5}$, and $P^{\Delta}_1 (I,T,p) = E^{\Delta}_1 (I,T,p) = \frac{4}{3}$, violating Boundedness and Constant Productivity.

\item \label{A04} The negative results are demonstrated by the following example. Let $I = T = \{1,2,3\}$. Let $p_1 = (2,1,1)$, $p_2 = (1,2,1)$, and $p_3 = (3,1,1)$. Note that $p_2$ is a permutation of $p_1$. The indicated rules compensate as follows: 
\begin{align*}
IC(I, T, p) &= (1,2,3) \text{,} & SV(I,T,p) &= \left( \frac{3}{2} , 2 , \frac{5}{2} \right) \text{,} \\
P^{\Delta}(I, T, p) &= \left( \frac{6}{5} , \frac{12}{5} , \frac{12}{5} \right) \text{,} & E^{\Delta}(I, T, p) &= \left( \frac{4}{3} , \frac{7}{3} , \frac{7}{3} \right) \text{.}
\end{align*}

\item \label{A07} This can be shown with the help of the following lemma.
\begin{lemma} \label{lemma01}
For any $(I,T,p)$ where $|T| \geq |I|+1$, for any $i,j \not\in I$ ($i=j$ is allowed), for any $p_i,p_j \in \mathbb{R}_+^T$:
\begin{enumerate}[(i)]
\item \label{lemma01i} If $p_i \geqq p_j$, then $y(I \cup \{ i \} , T , (p,p_i)) \geq y(I \cup \{ j \} , T , (p,p_j))$.
\item \label{lemma01ii} If $p_i > p_j$, then $y(I \cup \{ i \} , T , (p,p_i)) > y(I \cup \{ j \} , T , (p,p_j))$.
\end{enumerate}
\end{lemma}
\begin{proof}
We show the proof for part \ref*{lemma01i}. Let $a \in A^* (I \cup \{ j \} , T , (p,p_j))$. Since $p_i \geqq p_j$, we must have $p_i^{a(j)} \geq p_j^{a(j)}$, and hence 
\[
\sum_{k \in I\cup \{ i \}} p^{a(k)}_k \geq \sum_{k \in I\cup \{ j \}} p^{a(k)}_k \text{.}
\]
But by definition, $y(I \cup \{ i \} , T , (p,p_i))$ is weakly larger than the left-hand side of this inequality while $y(I \cup \{ j \} , T , (p,p_j)$ is equal to the right-hand side.
\end{proof}
To show that $SV$ satisfies Order Preservation, fix $(I,T,p)$ and $i,j \in I$. Assume $p_i \geqq p_j$. Note that for any $J \subset I \setminus \{ i,j \}$, \autoref*{lemma01}\ref*{lemma01i} implies 
\[
y(J \cup \{i\},T,p_{J \cup \{i\}}) \geq y(J \cup \{j\},T,p_{J \cup \{j\}}) \text{,}
\]
and thus
\[
y(J \cup \{i\},T,p_{J \cup \{i\}}) - y(J,T,p_J) \geq y(J \cup \{j\},T,p_{J \cup \{j\}}) - y(J,T,p_J)
\]
and
\begin{multline*}
y(J \cup \{i,j\},T,p_{J \cup \{i,j\}}) - y(J \cup \{j\},T,p_{J \cup \{j\}})\\
 \geq y(J \cup \{i,j\},T,p_{J \cup \{i,j\}}) - y(J \cup \{i\},T,p_{J \cup \{i\}}) \text{.}
\end{multline*}
Since this holds for any $J \subset I \setminus \{ i,j \}$, we must have $SV_i (I,T,p) \geq SV_j (I,T,p)$.

\item \label{A06} This was shown in \autoref{ex04}.

\item \label{A12} To show that $SV$ satisfies Strict Order Preservation, simply adapt the argument in \autoref{A07} above by invoking \autoref{lemma01}\ref{lemma01ii} and switching weak inequalities with strict inequalities.

\item \label{A13} The negative result is demonstrated by the following example. Let $I = T = \{1,2\}$. Let $p_1 = (2,1)$ and $p_2 = (2,0)$. Then $p_1 \geq p_2$. But $R_1 (I,T,p) = R_2 (I,T,p) = \frac{3}{2}$ for $R = SV, P^{\max}, P^{\Delta}, E^{\Delta}$, which violates Strong Order Preservation.

\item \label{A15} Since Efficiency and Symmetry are satisfied, the negative result is implied by \autoref{thm01}.

\item \label{A18} To show that $SV$ satisfies Individual Productivity Monotonicity, fix $(I,T,p)$ and $i \in I$. Let $\hat{p}_i$ satisfy $p_i \geq \hat{p}_i$. Note that for any $J \subset I \setminus \{ i \}$, \autoref{lemma01}\ref{lemma01i} implies
\[
y(J \cup \{i\},T,p_{J \cup \{i\}}) \geq y(J \cup \{i\},T,(p_{J},\hat{p}_i)) \text{,}
\]
and thus
\[
y(J \cup \{i\},T,p_{J \cup \{i\}}) - y(J,T,p_J) \geq y(J \cup \{i\},T,(p_{J},\hat{p}_i)) - y(J,T,p_J) \text{.}
\]
Since this holds for any $J \subset I \setminus \{ i \}$, we must have $SV_i (I,T,p) \geq SV_i (I,T,(p_{-i},\hat{p}_i))$.

\item \label{A17} This was shown in \autoref{ex05}.

\item \label{A21A} The negative result is demonstrated by the following example. Let $I = T = \{1,2\}$. Let $p_1 =(186 , 110)$, $\hat{p}_1 =(185 , 100)$, and $p_2 = (100 , 0)$. Then $p_1 > \hat{p}_1$, $P^{\Delta}_1 (I,T,p) \approx 172.39$, and $P^{\Delta}_1 (I,T,(\hat{p}_1 , p_2)) \approx 173.91$.

\item \label{A21} Note that for any $(I,T,p)$ and $i \in I$,
\[
E^{\Delta}_i (I,T,p) = \frac{1}{|I|} \left[ y(I,T,p) + \sum_{j \neq i} y(I \setminus \{j\},T,p_{-j}) \right] - \frac{|I|-1}{|I|}y(I \setminus \{i\},T,p_{-i}) \text{.}
\]

Now fix $(I,T,p)$ and $i \in I$. Let $\hat{p}_i$ satisfy $p_i \geq \hat{p}_i$. Set $\hat{p} = (p_{-i},\hat{p}_i)$. Note that \autoref{lemma01}\ref{lemma01i} implies
\[
y(I,T,p) \geq y(I,T,\hat{p})
\]
and
\[
y(I \setminus \{j\},T,p_{-j}) \geq y(I \setminus \{j\},T,\hat{p}_{-j})
\]
for every $j \in I$ where $j \neq i$. Since $p_{-i} = \hat{p}_{-i}$, we then get 
\begin{align*}
E^{\Delta}_i (I,T,p) &= \frac{1}{|I|} \left[ y(I,T,p) + \sum_{j \neq i} y(I \setminus \{j\},T,p_{-j}) \right] - \frac{|I|-1}{|I|}y(I \setminus \{i\},T,p_{-i}) \\
&\geq \frac{1}{|I|} \left[ y(I,T,\hat{p}) + \sum_{j \neq i} y(I \setminus \{j\},T,\hat{p}_{-j}) \right] - \frac{|I|-1}{|I|}y(I \setminus \{i\},T,\hat{p}_{-i}) \\
&= E^{\Delta}_i (I,T,\hat{p}) \text{.}
\end{align*}

\item \label{A22} To show that $SV$ satisfies Strict Individual Productivity Monotonicity, simply adapt the argument in \autoref{A18} above by invoking \autoref{lemma01}\ref{lemma01ii} and switching weak inequalities with strict inequalities.

\item \label{A22a} To show that $E^\Delta$ satisfies Strict Individual Productivity Monotonicity, simply adapt the argument in \autoref{A21} above by invoking \autoref{lemma01}\ref{lemma01ii} and switching weak inequalities with strict inequalities.

\item \label{A23} Continuing the example from \autoref{A13}, let $\hat{p}_2 = (2,1)$. Then $\hat{p}_2 \geq p_2$, yet $R_2 (I,T,p) = R_2 (I,T,(p_1,\hat{p}_2)) = \frac{3}{2}$ for $R =  SV, P^{\max}, P^{\Delta}, E^{\Delta}$, violating Strong Individual Productivity Monotonicity.

\item \label{A30} Since Efficiency is satisfied, the negative result is implied by \autoref{thm03}.

\item \label{A28} Since Efficiency is satisfied, the negative result is implied by \autoref{thm02}.

\item \label{A29} Before showing that $IC$ satisfies Consistency, we introduce some notation and establish some preliminary results.

Let $I,T \in \mathcal{N}$ satisfy $|I| \leq |T|$. For any assignment $a \in A (I,T)$ and for any $I' \subset I$, let $a|_{I'}$ denote the restriction of $a$ to $I'$. In addition, let  $I' \subset I$ and $T' \subset T$ satisfy $|I'| \leq |T'|$ and $|I \setminus I'| \leq |T \setminus T'|$. Then for $a' \in A(I',T')$ and $a \in A(I \setminus I' , T \setminus T')$, let $a'I'a$ denote the assignment in $A(I,T)$ that assigns according to $a'$ on $I'$ and according to $a$ on $I \setminus I'$.

Fix $(I,T,p)$ and $I' \subset I$. Define 
\[
\mathcal{T} \coloneqq \left\{ T' \subset T : \exists a \in A^* (I,T,p) \text{ such that } a(I') = T' \right\} \text{.}
\]
For every $T' \in \mathcal{T}$, define
\[
B(T') \coloneqq \left\{ a \in A^* (I,T,p) : a(I') = T' \right\} \text{,}
\]
which must be nonempty by definition of $\mathcal{T}$.

\begin{lemma} \label{lemma14}
For every $T' \in \mathcal{T}$,
\[
B(T') = \left\{ a'I'a \in A (I,T) : a' \in A^* (I', T' ,p_{I'}^{T'}) \text{ and } a \in A^* ( I \setminus I' , T \setminus T' , p_{I \setminus I'}^{T \setminus T'}) \right\} \text{.}
\]
\end{lemma}
\begin{proof}
Fix $T' \in \mathcal{T}$.

Choose $a \in B(T')$. Then by definition, $a(I') = T'$ and $a \in A^*(I,T,p)$. The former implies $a|_{I'} \in A(I',T')$, while the latter implies $\sum_{i \in I} p_{i}^{a(i)} \geq \sum_{i \in I} p_{i}^{\hat{a}(i)}$ for every $\hat{a} \in A(I,T)$. In particular, we must have
\[
\sum_{i \in I'} p_{i}^{a(i)} + \sum_{i \in I \setminus I'} p_{i}^{a(i)} \geq \sum_{i \in I'} p_{i}^{\hat{a}(i)} +\sum_{i \in I \setminus I'} p_{i}^{a(i)}
\]
for every $\hat{a} \in A(I',T')$. But then this implies $\sum_{i \in I'} p_{i}^{a(i)}  \geq \sum_{i \in I'} p_{i}^{\hat{a}(i)}$ for every $\hat{a} \in A(I',T')$, or $a|_{I'} \in A^*(I',T', p_{I'}^{T'})$. Similar reasoning can show $a|_{I \setminus I'} \in A^* ( I \setminus I' , T \setminus T' , p_{I \setminus I'}^{T \setminus T'})$.

Going the other direction, choose $a' \in A^* (I', T' ,p_{I'}^{T'})$ and $a \in A^* ( I \setminus I' , T \setminus T' , p_{I \setminus I'}^{T \setminus T'})$. Since $B(T')$ is nonempty, choose $a^* \in B(T')$. Then we have
\[
\sum_{i \in I'} p_{i}^{a'(i)} \geq \sum_{i \in I'} p_{i}^{a^*(i)}
\]
and
\[
\sum_{i \in I \setminus I'} p_{i}^{a(i)} \geq \sum_{i \in I \setminus I'} p_{i}^{a^*(i)} \text{.}
\]
But then
\[
\sum_{i \in I'} p_{i}^{a'(i)} + \sum_{i \in I \setminus I'} p_{i}^{a(i)} \geq \sum_{i \in I} p_{i}^{a^*(i)} \text{,}
\]
which implies $a'I'a \in A^*(I,T,p)$ since $a^* \in A^*(I,T,p)$. Finally, note that $a'I'a(I') = a'(I') = T'$. Thus, we have shown $a'I'a \in B(T')$.
\end{proof}

For every $T' \in \mathcal{T}$ and for every $a' \in A^* (I',T',p_{I'}^{T'})$, define
\[
C(a';T') \coloneqq \left\{ a \in A^* (I,T,p) : a|_{I'} = a'  \right\} \text{.}
\]
For fixed $T' \in \mathcal{T}$, \autoref{lemma14} implies $A^* (I',T',p_{I'}^{T'})$ is nonempty, and that $C(a';T')$ is nonempty for every $a' \in A^* (I',T',p_{I'}^{T'})$.

\begin{lemma} \label{lemma15}
Fix $T' \in \mathcal{T}$. For every $a' \in A^* (I',T',p_{I'}^{T'})$, 
\[
|C(a' ; T')| = |A^* ( I \setminus I' , T \setminus T' , p_{I \setminus I'}^{T \setminus T'})| \text{.}
\]
\end{lemma}
\begin{proof}
Fix $T' \in \mathcal{T}$. Choose $a' \in A^* (I',T',p_{I'}^{T'})$. We show that there is a bijection from $C(a' ; T')$ to $A^* ( I \setminus I' , T \setminus T' , p_{I \setminus I'}^{T \setminus T'})$. Namely, $a \mapsto a|_{I \setminus I'}$.

First, for $a \in C(a' ; T')$, \autoref{lemma14} implies $a'|_{I \setminus I'} \in A^* ( I \setminus I' , T \setminus T' , p_{I \setminus I'}^{T \setminus T'})$. Next, if $a , \hat{a} \in C(a' ; T')$ satisfy $a|_{I \setminus I'} = \hat{a}|_{I \setminus I'}$, then since $a' = a|_{I'} = \hat{a}|_{I'}$, we must have $a = \hat{a}$. Finally, choose $\hat{a}' \in A^* ( I \setminus I' , T \setminus T' , p_{I \setminus I'}^{T \setminus T'})$. Then \autoref{lemma14} implies that $a'I'\hat{a}' \in B(T')$ and therefore $a'I'\hat{a}' \in A^*(I,T,p)$. Since $a'I'\hat{a}'|_{I'} = a'$, we have established $a'I'\hat{a}' \in C(a' ; T')$.
\end{proof}

Now we are ready to show that $IC$ satisfies Consistency. Fix $i \in I'$. It is sufficient to show
\begin{equation} \label{eq04}
\sum_{a \in A^*(I,T,p)} p_i^{a(i)} = \sum_{a \in A^*(I,T,p)} \frac{1}{\left| A^*\left( I' , a(I') , p_{I'}^{a(I')} \right) \right| } \sum_{\hat{a} \in A^* \left( I' , a(I') , p_{I'}^{a(I')} \right) }  p_i^{\hat{a}(i)} \text{.}
\end{equation}

Note that 
\[
\mathcal{B} \coloneqq \left\{ B(T') : T' \in \mathcal{T} \right\}
\]
is a partition of $A^* (I,T,p)$. In addition, if $T',T'' \in \mathcal{T}$ are distinct, then $B(T') \neq B(T'')$. Thus $T' \mapsto B(T')$ defines a bijection from $\mathcal{T}$ to $\mathcal{B}$. Also, note that by definition, $( I' , a'(I') , p_{I'}^{a'(I')} ) = ( I' , \hat{a}'(I') , p_{I'}^{\hat{a}'(I')} )$ whenever $a'$ and $\hat{a}'$ are in the same cell of this partition.  Thus the right-hand side of \autoref*{eq04} is equal to 
\begin{align} 
\sum_{a \in A^*(I,T,p)} & \frac{1}{\left| A^*\left( I' , a(I') , p_{I'}^{a(I')} \right) \right| } \sum_{a' \in A^* \left( I' , a(I') , p_{I'}^{a(I')} \right) }  p_i^{a'(i)} \nonumber \\
&= \sum_{B \in \mathcal{B}} \sum_{a \in B}  \frac{1}{\left| A^*\left( I' , a(I') , p_{I'}^{a(I')} \right) \right|} \sum_{a' \in A^* \left( I' , a(I') , p_{I'}^{a(I')} \right) }  p_i^{a'(i)} \nonumber \\
&= \sum_{T' \in \mathcal{T}}  \frac{|B(T')|}{\left| A^*\left( I' , T' , p_{I'}^{T'} \right) \right| } \sum_{a' \in A^* \left( I' , T' , p_{I'}^{T'} \right) }  p_i^{a'(i)} \label{eq03} \text{.}
\end{align}

Next, note that 
\[
\mathcal{C}(T') \coloneqq \left\{ C(a' ; T') : a' \in A^* (I',T',p_{I'}^{T'}) \right\}
\]
is a partition of $B(T')$. In addition, if $a',\hat{a}' \in A^* (I',T',p_{I'}^{T'})$ are distinct, then $C(a' ; T') \neq C(\hat{a}' ; T')$. Thus $a' \mapsto C(a' ; T')$ defines a bijection from $A^* (I',T',p_{I'}^{T'})$ to $\mathcal{C}(T')$.
Thus the left-hand side of \autoref*{eq04} is equal to 
\begin{align}
\sum_{a \in A^*(I,T,p)} p_i^{a(i)} &= \sum_{T' \in \mathcal{T}} \sum_{a' \in A^* (I',T',p_{I'}^{T'})} \sum_{a \in C(a' ; T')} p_i^{a(i)} \nonumber \\
&= \sum_{T' \in \mathcal{T}} \sum_{a' \in A^* (I',T',p_{I'}^{T'})} |C(a' ; T')| \, p_i^{a'(i)} \label{eq05} \text{,}
\end{align}
where the last equality follows from the fact that $p_i^{a(i)} = p_i^{a'(i)}$ for every $a \in C(a' ; T')$.

\autoref{lemma14} implies
\[
|B(T')| = |A^* (I', T' ,p_{I'}^{T'})| \times |A^* ( I \setminus I' , T \setminus T' , p_{I \setminus I'}^{T \setminus T'})|\text{,}
\]
while \autoref{lemma15} asserts
\[
|C(a' ; T')| = |A^* ( I \setminus I' , T \setminus T' , p_{I \setminus I'}^{T \setminus T'})| \text{.}
\]
This implies \autoref*{eq04}, as desired.

\item \label{A34} The following lemma will be useful. (The proof is omitted as it is obvious.)
\begin{lemma} \label{lemma09}
For any problem $(I,T,p)$, if $i \in I$ is a null worker, then 
\[
y(I , T , p ) = y(I \setminus \{ i \} , T , p_{-i } ) \text{.}
\]
\end{lemma}

Now fix problem $(I,T,p)$ and let $i \in I$ be a null worker. Fix $j \in I \setminus \{ i \}$. Define the following sets:
\begin{align*}
\mathcal{I} &= \left\{ J : J \subset I \setminus \{ j \} \right\} \text{,} \\
\mathcal{I}_{-i} &= \left\{ J : J \subset I \setminus \{ i,j \} \right\} \text{, and } \\
\mathcal{I}_{+i} &= \left\{ J : J \in \mathcal{I} \text{ and } i \in J \right\} \text{.}
\end{align*}
Therefore, $\mathcal{I} = \mathcal{I}_{-i} \cup \mathcal{I}_{+i}$,  
\[
SV_j (I,T,p) = \sum_{J \in \mathcal{I}} \frac{|J|!(|I| - |J|-1)!}{|I|!} \left[ y(J \cup \{j\},T,p_{J \cup \{j\}}) - y(J,T,p_J) \right] \text{,}
\]
and 
\[
SV_j (I\setminus \{ i \} , T , p_{-i }) = \sum_{J \in \mathcal{I}_{-i}} \frac{|J|!(|I| - |J|-2)!}{(|I| - 1)!} \left[ y(J \cup \{j\},T,p_{J \cup \{j\}}) - y(J,T,p_J) \right] \text{.}
\]
To establish the result, we must show $SV_j (I,T,p) = SV_j (I\setminus \{ i \} , T , p_{-i })$.

Note that
\begin{align*}
SV_j (I,T,p) &= \sum_{J \in \mathcal{I}_{-i}} \frac{|J|!(|I| - |J|-1)!}{|I|!} \left[ y(J \cup \{j\},T,p_{J \cup \{j\}}) - y(J,T,p_J) \right] \\
& \quad + \sum_{J' \in \mathcal{I}_{+i}} \frac{|J'|!(|I| - |J'|-1)!}{|I|!} \left[ y(J' \cup \{j\},T,p_{J' \cup \{j\}}) - y(J',T,p_{J'}) \right]  \text{.}
\end{align*}
There is a bijective function from $\mathcal{I}_{-i}$ to $\mathcal{I}_{+i}$, namely $J \mapsto J \cup \{ i \}$. Since $|J \cup \{i \}| = |J|+1$ for $J \in \mathcal{I}_{-i}$, the above equation becomes
\begin{align*}
SV_j (I,T,p) &= \sum_{J \in \mathcal{I}_{-i}} \left( \frac{|J|!(|I| - |J|-1)!}{|I|!} \left[ y(J \cup \{j\},T,p_{J \cup \{j\}}) - y(J,T,p_J) \right] \right. \\
& \left. \quad\quad + \frac{(|J|+1)!(|I| - |J|-2)!}{|I|!} \left[ y(J \cup \{i,j\},T,p_{J \cup \{i,j\}}) - y(J \cup \{ i \} , T , p_{J \cup \{i\} } ) \right]  \right) \text{.}
\end{align*}
By \autoref{lemma09}, $y(J \cup \{i,j\},T,p_{J \cup \{i,j\}}) = y(J \cup \{j\},T,p_{J \cup \{j\}})$ and $y(J \cup \{i\},T,p_{J \cup \{i\}}) = y(J ,T,p_{J})$. Thus we have
\begin{align*}
SV_j (I,T,p) &= \sum_{J \in \mathcal{I}_{-i}} \left( \frac{|J|!(|I| - |J|-1)!}{|I|!} + \frac{(|J|+1)!(|I| - |J|-2)!}{|I|!} \right) \\
& \quad\quad \times \left[ y(J \cup \{j\},T,p_{J \cup \{j\}}) - y(J,T,p_J) \right] \\
&= \sum_{J \in \mathcal{I}_{-i}} \frac{|J|!(|I| - |J|-2)!}{(|I| - 1)!}  \left[ y(J \cup \{j\},T,p_{J \cup \{j\}}) - y(J,T,p_J) \right] \\
&= SV_j (I\setminus \{ i \} , T , p_{-i }) \text{.}
\end{align*}

\item \label{A33} Fix problem $(I,T,p)$. Let $i \in I$ be a null worker. Set $I' = I \setminus \{i\}$ and $M = |T|-|I|+1$. Then for $a' \in A(I',T)$ and $t \in T \setminus a'(I')$, let $a'I't$ denote the assignment in $A(I,T)$ that assigns according to $a'$ on $I'$ and assigns $t$ to $i$.

\begin{lemma} \label{lemma07}
$A^*(I,T,p) = \left\{ a'I't : a' \in A^*(I',T,p_{-i}), t \in T \setminus a'(I') \right\}$.
\end{lemma}

The proof is omitted as it is straightforward. One implication of \autoref{lemma07} is the following.
\begin{align*}
|A^*(I,T,p)| &= |\left\{ a'I't : a' \in A^*(I',T,p_{-i}), t \in T \setminus a'(I') \right\}| \\
&= |A^*(I',T,p_{-i})|(|T|-|I'|) \\
&= |A^*(I',T,p_{-i})| M
\end{align*}
Thus for $j \in I'$, we have
\begin{align*}
IC_j (I  , T , p ) &= \frac{ 1 }{| A^* (I, T, p) |} \sum_{a \in A^*(I,T,p)} p_j^{a(j)} \\
&= \frac{ 1 }{| A^* (I', T, p_{-i}) |M} \sum_{a' \in A^*(I',T,p_{-i})} \sum_{t \in T \setminus a(I')} p_j^{a'I't(j)} \\
&= \frac{ 1 }{| A^* (I', T, p_{-i}) |M} \sum_{a' \in A^*(I',T,p_{-i})} M p_j^{a'(j)} \\
&= \frac{ 1 }{| A^* (I', T, p_{-i}) |} \sum_{a' \in A^*(I',T,p_{-i})}  p_j^{a'(j)} \\
&= IC_j (I' , T , p_{-i} ) \text{.}
\end{align*}

\item \label{A36} Since Efficiency is satisfied, the negative result is implied by \autoref{prop01}.

\item \label{A39} We prove the result for the family of Parametric rules (of which $E$, $P^{Av}$, and $P^{\max}$ are members).

Let $Par^f$ be a Parametric rule. Fix the problem $(I,T,p)$. Fix $I' \subset I$. Let $\lambda$ satisfy $Par^f_i (I,T,p) = f(p_i , \lambda)$ for every $i \in I$. Let $\lambda'$ satisfy $Par^f_i (I',T,p_{I'}) = f(p_i , \lambda')$ for every $i \in I'$. Either $\lambda \leq \lambda'$ or $\lambda \geq \lambda'$. If the former, then $Par^f_{I'} (I,T,p) \leqq Par^{f} (I',T,p_{I'})$ since $f$ is weakly monotone in $\lambda$. If the latter, then $Par_{I'}^{f} (I,T,p) \geqq Par^{f} (I',T,p_{I'})$.

\item \label{A38} Fix the problem $(I,T,p)$. Fix $I' \subset I$ and $i \in I'$. Define the following sets:
\begin{align*}
\hat{I} &= I \setminus I' \text{,} \\
\mathcal{I} &= \left\{ J : J \subset I \setminus \{ i \} \right\} \text{,} \\
\mathcal{I}' &= \left\{ J : J \subset I' \setminus \{ i \} \right\} \text{, and} \\
\hat{\mathcal{I}} &= \left\{ J : J \subset \hat{I} \right\} \text{.}
\end{align*}
Then
\begin{align*}
SV_i (I,T,p) &= \sum_{J \in \mathcal{I}} \frac{|J|!(|I| - |J|-1)!}{|I|!} \left[ y(J \cup \{i\},T,p_{J \cup \{i\}}) - y(J,T,p_J) \right] \\
&= \sum_{J' \in \mathcal{I}'}  \sum_{\hat{J} \in \hat{\mathcal{I}}} \frac{(|J'| + | \hat{J}|)!(|I| - |J'| - |\hat{J}| -1)!}{|I|!}  \\
& \quad\quad \times  \left[ y(J' \cup \hat{J} \cup  \{i\},T,p_{J' \cup \hat{J} \cup \{i\}}) - y( J' \cup \hat{J} , T,p_{J' \cup \hat{J}}) \right] 
\end{align*}
Because $y$ is submodular (see \autoref{Sec_NHfH}), this implies
\begin{align}
SV_i (I,T,p) &\leq  \sum_{J' \in \mathcal{I}'} \left[ y(J' \cup  \{i\},T,p_{J' \cup \{i\}}) - y( J'  , T,p_{J'}) \right] \nonumber \\
& \quad\quad \times \sum_{\hat{J} \in \hat{\mathcal{I}}} \frac{(|J'| + | \hat{J}|)!(|I| - |J'| - |\hat{J}| -1)!}{|I|!} \label{eq02}
\end{align}
For fixed $J' \in \mathcal{I}'$,
\begin{align*}
\sum_{\hat{J} \in \hat{\mathcal{I}}} \frac{(|J'| + | \hat{J}|)!(|I| - |J'| - |\hat{J}| -1)!}{|I|!} &=  \sum_{n=0}^{|\hat{I}|} \genfrac(){0pt}{0}{|\hat{I}|}{n} \frac{(|J'| + n)!(|I| - |J'| - n -1)!}{|I|!} \\
&= \frac{|\hat{I}|!}{|I|!} \sum_{n=0}^{|\hat{I}|}  \frac{(|J'| + n)!(|I| - |J'| - n -1)!}{n!(|\hat{I}| - n)!} \\
&= |J'|! (|I'| - |J'|  -1)! \frac{|\hat{I}|!}{|I|!} \\
& \quad\quad \times \sum_{n=0}^{|\hat{I}|} \genfrac(){0pt}{0}{|J'| + n}{n} \genfrac(){0pt}{0}{|I| - |J'| - n -1}{|\hat{I}| - n}  
\end{align*}
A generalization of Vandermonde's convolution \citep[see][]{gould1956sgo} implies 
\begin{align*}
\sum_{n=0}^{|\hat{I}|} \genfrac(){0pt}{0}{|J'| + n}{n} \genfrac(){0pt}{0}{|I| - |J'| - n -1}{|\hat{I}| - n}   = \genfrac(){0pt}{0}{|I|}{|\hat{I}|}
\end{align*}
Thus we have
\begin{align*}
\sum_{\hat{J} \in \hat{\mathcal{I}}} \frac{(|J'| + | \hat{J}|)!(|I| - |J'| - |\hat{J}| -1)!}{|I|!} &=  \frac{|J'|! (|I'| - |J'|  -1)!}{|I'|} \text{,}
\end{align*}
which means that \autoref*{eq02} implies 
\begin{align*}
SV_i (I,T,p) &\leq  \sum_{J' \in \mathcal{I}'} \frac{|J'|! (|I'| - |J'|  -1)!}{|I'|} \left[ y(J' \cup  \{i\},T,p_{J' \cup \{i\}}) - y( J'  , T,p_{J'}) \right]  \\
&= SV_i (I',T,p_{I'}) \text{.}
\end{align*}

\item \label{A37} This was shown in \autoref{ex06}.

\item \label{A40} Continuing the example from \autoref{A02b}, note $P^{\Delta} (I,T,p) = E^{\Delta} (I,T,p) =( \frac{4}{3} , \frac{4}{3} )$.  Let $I' = \{1,2\}$. Then $P^{\Delta}_1 (I',T,p_{I'}) = E^{\Delta}_1 (I,T,p) = 1$ and $P^{\Delta}_2 (I',T,p_{I'}) = E^{\Delta}_2 (I,T,p) = 2$, which violates Solidarity in Hiring.

\item \label{A42} This was shown in \autoref{ex07}.

\item \label{A47} Note that for any problem $(I,T,p)$ and for every $T'$ satisfying $T^*(I,T,p) \subset T' \subset T$, we must have $A^* (I,T,p) = A^* (I,T',p^{T'})$. The result follows easily from this fact.

\item \label{A49} The negative result is demonstrated by the following example. Let $I = \{1,2\}$ and $T = \{1,2,3\}$. Let $p_1 = (2,0,1)$ and $p_2 = (0,1,0)$. Then $P^{Av} (I,T,p) = (\frac{9}{4} , \frac{3}{4})$. But for $T' = T^* (I,T,p) = \{ 1,2 \}$, we have $P^{Av} (I,T',p^{T'}) = (2 , 1)$.

\item \label{A51} The result follows easily from the following lemma. 
\begin{lemma} \label{lemma12}
For every $I,T \subset \mathcal{N}$ where $|I| \leq |T|$, for every $p,\hat{p} \in \mathbb{R}_{+}^{I \times T}$ satisfying $A^* (I,T,p) \cap A^* (I,T,\hat{p}) \neq \emptyset$, we have $y(I,T,p+\hat{p}) = y(I,T,p) + y(I,T,\hat{p})$ and $A^* (I,T,p+ \hat{p}) = A^* (I,T,p) \cap A^* (I,T,\hat{p})$.
\end{lemma}
\begin{proof}
Let $I,T \subset \mathcal{N}$ satisfy $|I| \leq |T|$. Let $p,\hat{p} \in \mathbb{R}_{+}^{I \times T}$ satisfy $A^* (I,T,p) \cap A^* (I,T,\hat{p}) \neq \emptyset$. Fix $a \in A^* (I,T,p) \cap A^* (I,T,\hat{p})$ and $\hat{a} \in A^* (I,T,p+ \hat{p})$. Then by definition, we must have
\[
\sum_{i \in I} (p+\hat{p})^{\hat{a}(i)}_i \geq \sum_{i \in I} (p+\hat{p})^{a(i)}_i
\]
and
\[
\sum_{i \in I} p^{\hat{a}(i)}_i + \sum_{i \in I} \hat{p}^{\hat{a}(i)}_i \leq \sum_{i \in I} p^{a(i)}_i + \sum_{i \in I} \hat{p}^{a(i)}_i \text{.}
\]
Note that for each of the above inequalities, the left-hand sides are equal and the right-hand sides are equal. Thus 
\[
\sum_{i \in I} (p+\hat{p})^{\hat{a}(i)}_i = \sum_{i \in I} p^{a(i)}_i + \sum_{i \in I} \hat{p}^{a(i)}_i \text{,}
\]
\[
\sum_{i \in I} (p+\hat{p})^{\hat{a}(i)}_i = \sum_{i \in I} (p+\hat{p})^{a(i)}_i \text{,}
\]
and
\[
\sum_{i \in I} p^{\hat{a}(i)}_i + \sum_{i \in I} \hat{p}^{\hat{a}(i)}_i = \sum_{i \in I} p^{a(i)}_i + \sum_{i \in I} \hat{p}^{a(i)}_i \text{.}
\]
The first equation implies $y(I,T,p+\hat{p}) = y(I,T,p) + y(I,T,\hat{p})$. The second equation implies $A^* (I,T,p+ \hat{p}) \supset A^* (I,T,p) \cap A^* (I,T,\hat{p})$. The third equation implies $A^* (I,T,p+ \hat{p}) \subset A^* (I,T,p) \cap A^* (I,T,\hat{p})$.
\end{proof}

\item \label{A53} This was shown in \autoref{ex09}.

\item \label{A52} This was shown in \autoref{ex08}.

\item \label{A55} Continuing \autoref{ex09}, note that
\begin{align*}
P^{\max} (I,T,p) &= (\tfrac{12}{5} , \tfrac{8}{5}) \text{,} &  P^{\Delta} (I,T,p) &= (\tfrac{8}{3} , \tfrac{4}{3}) \text{,} &  E^{\Delta} (I,T,p) &= (\tfrac{5}{2} , \tfrac{3}{2}) \text{,}
\end{align*} 
while 
\[
P^{\max} (I,T,\hat{p}) = P^{\Delta} (I,T,\hat{p}) = E^{\Delta} (I,T,\hat{p}) = (1,1)
\]
and 
\[
P^{\max} (I,T,p+\hat{p}) = P^{\Delta} (I,T,p+\hat{p}) = E^{\Delta} (I,T,p+\hat{p}) = (4,2) \text{.}
\]

\item \label{A56} Let $I,T \subset \mathcal{N}$ satisfy $|I| \leq |T|$, and let $p,\hat{p} \in \mathbb{R}_{+}^{I \times T}$ satisfy $A^* (I,T,p) = A^* (I,T,\hat{p})$. \autoref{lemma12} implies $A^* (I,T,p) = A^* (I,T,\hat{p}) = A^* (I,T,p+\hat{p})$. Then for $i \in I$, 
\begin{align*}
IC_i (I,T,p+\hat{p}) &= \frac{ 1 }{|A^* (I,T,p+\hat{p})|} \sum_{a \in A^* (I,T,p+\hat{p})} (p+\hat{p})_{i}^{a(i)} \\
&= \frac{ 1 }{|A^* (I,T,p)|} \sum_{a \in A^* (I,T,p)} p_{i}^{a(i)} + \frac{ 1 }{|A^* (I,T,\hat{p})|} \sum_{a \in A^* (I,T,\hat{p})} \hat{p}_{i}^{a(i)} \\
&= IC_i (I,T,p) + IC_i (I,T,\hat{p}) \text{.}
\end{align*}

\end{enumerate}

\section{Weak Consistency and Weak Individual Contribution Rules} \label{Sec_App_IC}

In this appendix, we further explore the axiom Weak Consistency. We show that Weak Consistency and Efficiency almost characterize a family of compensation rules that we call the Weak Individual Contribution rules. $IC$ is a member of this family, but we introduce others. We also show that the set of compensation problems with multiple optimal assignments is ``nowhere dense'' --- a topological property that captures the idea that the set is sparse. Since the Weak Individual Contribution rules potentially differ only in how they award compensation for problems with multiple optimal assignments, this result shows that Weak Individual Contribution rules award the same compensation for almost all problems.

We say a rule $R$ is a \emph{Weak Individual Contribution rule} if for any problem $(I,T,p)$ where $A^*(I,T,p) = \{a\}$ is a singleton, and for any $i \in I$, 
\[
R_i (I,T,p) =  p_{i}^{a(i)} \text{.}
\]

Weak Individual Contribution rules only differ in how they compensate when there is not a unique optimal assignment of tasks. $IC$ is obviously a member of this family. 

Another Weak Individual Contribution rule is one that assigns a priority ranking to the set of workers and then uses that ranking to assign compensation when there are multiple optimal assignments.

\begin{axiom}[Individual Contribution Rule with Priority $\prec$] 
Let $\prec$ be a strict ordering over $\mathbb{N}$. For problem $(I,T,p)$, let $i^{\min}$ and $i^{\max}$ denote the first and last workers in $I$ with respect to $\prec$, respectively. Set
\[
A^{i^{\min}} \coloneqq \{ a \in A^* (I,T,p) : p_{i^{\min}}^{a(i^{\min})} \geq p_{i^{\min}}^{\hat{a}(i^{\min})} \text{ for every } \hat{a} \in A^*(I,T,p) \} \text{.}
\]
For $i \neq i^{\max}$, let $s(i)$ denote the immediate successor of $i$ in $I$ with respect to $\prec$, and recursively define
\[
A^{s(i)} \coloneqq \{ a \in A^i (I,T,p) : p_{s(i)}^{a(s(i))} \geq p_{s(i)}^{\hat{a}(s(i))} \text{ for every } \hat{a} \in A^i(I,T,p) \} \text{.}
\]
Note that $A^{i^{\max}}$ is non-empty, and while it may not be a singleton, every $a \in A^{i^{\max}}$ yields the same output levels for every worker. I.e. $p_i^{a(i)} = p_i^{\hat{a}(i)}$ for every $i \in I$ and every $a,\hat{a} \in A^{i^{\max}}$.

For any problem $(I,T,p)$ and for any $i \in I$, the Individual Contribution with Priority $\prec$ rule $IC^{\prec}$ assigns to $i$ the compensation
\[
IC^{\prec}_i (I,T,p) \coloneqq p_{i}^{a(i)} \text{,}
\]
for $a \in A^{i^{\max}}$.
\end{axiom}

The next rule is similar, but the method for choosing the optimal assignment is not based on an ordering of the workers, but instead depends only on the set of optimal assignments.

\begin{axiom}[Individual Contribution Rule with Choice $c$] 
Let $c$ be a choice function defined over all sets of possible assignments. For any problem $(I,T,p)$ and for any $i \in I$, $IC^{c}$ assigns to $i$ the compensation
\[
IC_i^{c} (I,T,p) \coloneqq p_{i}^{a(i)} \text{,}
\]
where $a = c(A^*(I,T,p))$.
\end{axiom}

\begin{proposition} \label{prop02}
Every Weak Individual Contribution rule satisfies Weak Consistency.
\end{proposition}
\begin{proof}
This result follows easily from the following fact: For any problem $(I,T,p)$ where $A^* (I,T,p) = \{ a \}$, and for any $I' \subset I$, we have  $A^* (I',a(I'),p_{I'}^{a(I')}) = \{ a' \}$, where $a'$ is the restriction of $a$ to $I'$.
\end{proof}

\begin{proposition} \label{prop03}
Every rule satisfying Efficiency and Weak Consistency is a Weak Individual Contribution rule.
\end{proposition}
\begin{proof}
Let $(I,T,p)$ satisfy $A^* (I,T,p) = \{ a \}$. Fix $i \in I$. By Weak Consistency, 
\[
R_i (I,T,p) = R(\{i\},a(i),p_i^{a(i)}) \text{.}
\]
Note that $(\{i\},a(i),p_i^{a(i)})$ is a problem with one individual, so Efficiency implies $R(\{i\},a(i),p_i^{a(i)}) = p_i^{a(i)}$, which proves the result.
\end{proof}

Together, Propositions \ref*{prop02} and \ref*{prop03} almost characterize the Weak Individual Contribution rules; the only gap is that, in general, the Weak Individual Contribution rules are not Efficient. However, these propositions show that Weak Consistency is the defining property of these rules.

All Weak Individual Contribution rules agree when the compensation problem has a unique optimal assignment, and thus disagree only when a problem has multiple optimal assignments. Our final result shows that such problems are relatively rare.

We say a set $Y$ is \emph{nowhere dense} in the topological space $X$ if there is no open set $U$ of $X$ in which $Y \cap U$ is dense in $U$.

\begin{lemma}
Fix $I,T \in \mathcal{N}$ where $|I| \leq |T|$. Endow $\mathbb{R}_+^{I \times T}$ with the Euclidean topology. Then the set
\[
M \coloneqq \left\{ p \in \mathbb{R}_+^{I \times T} : |A^*(I,T,p)| > 1 \right\}
\]
is nowhere dense in $\mathbb{R}_+^{I \times T}$.
\end{lemma}
\begin{proof}
Define
\[
S \coloneqq \left\{ p \in \mathbb{R}_+^{I \times T} : |A^*(I,T,p)| = 1 \right\} \text{.}
\]
Then obviously $S$ and $M$ form a bipartition of $\mathbb{R}_+^{I \times T}$. It is straightforward to show that $S$ is open, and therefore $M$ is closed. Thus $M$ is nowhere dense if and only if $S$ is dense in $\mathbb{R}_+^{I \times T}$.

So let $U$ be an open set in $\mathbb{R}_+^{I \times T}$. Choose $p \in U$ and $a \in A^*(I,T,p)$. For any $\epsilon > 0$, define the productivity matrix $p(\epsilon)$ as follows. For any $i \in I$ and $t \in T$,
\[
p(\epsilon)_i^t \coloneqq \begin{cases}
p_i^t + \epsilon & \text{ if } t = a(i) \\
p_i^t & \text{ otherwise.}
\end{cases}
\]
Since $U$ is open, there obviously exists $\epsilon >0$ such that $p(\epsilon) \in U$.

We claim that $|A^*(I,T,p(\epsilon)|=1$, thus showing that $p(\epsilon) \in U \cap S$. Since $a \in A^*(I,T,p)$, we have $\sum_{i \in I} p_i^{a(i)} \geq \sum_{i \in I} p_i^{a'(i)}$ for every $a' \in A(I,T)$. Note also that
\[
\sum_{i \in I} p(\epsilon)_i^{a(i)} = \sum_{i \in I} p_i^{a(i)} + |I| \epsilon \text{,}
\]
while for any $a' \in A(I,T)$ such that $a' \neq a$, we have
\[
\sum_{i \in I} p(\epsilon)_i^{a'(i)} = \sum_{i \in I} p_i^{a'(i)} + |J| \epsilon
\]
where $J = \{ i \in I : a(i) = a'(i) \}$. When $a' \neq a$, we must have $|J| < |I|$ and thus 
\[
\sum_{i \in I} p(\epsilon)_i^{a(i)} > \sum_{i \in I} p(\epsilon)_i^{a'(i)} \text{.}
\]
Since this is true for every $a' \neq a$, we must have $\{a\} = A^*(I,T,p(\epsilon))$.

We have thus shown that $U \cap S$ is non-empty for any open $U$, and thus $S$ is dense in $\mathbb{R}_+^{I \times T}$.
\end{proof}

%%%%%
%%Bibliography
%%%%%

\clearpage

\bibliographystyle{te}
\bibliography{johnbib}

\end{document}